\documentclass[aps,prb,amsmath,amssymb,footinbib,showpacs,twocolumn]{revtex4-2}
\usepackage{amsmath}
\usepackage{amssymb}
\usepackage{amsthm}
\usepackage{setspace}
\usepackage{graphicx}
\usepackage{braket}
\usepackage{mathrsfs}
\usepackage{float}
\usepackage[colorlinks = true,linkcolor = blue,urlcolor  = blue,citecolor = blue,anchorcolor = blue]{hyperref}
\usepackage[utf8]{inputenc}
\usepackage[english]{babel}
\usepackage{bm}

\begin{document}

\title{Chiral anomaly and positive longitudinal magnetoresistance in the type-II Dirac semimetals $\it{A}_x$PdTe$_2$ (\textit{A} = Cu, Ag)}
\author{Sonika$^1$}
\author{Sunil Gangwar$^1$}
\author{Nikhlesh Singh Mehta$^2$}
\author{G. Sharma$^1$}
\author{C. S. Yadav$^1$}
\email{shekhar@iitmandi.ac.in}
\affiliation{$^1$School of Physical Sciences, Indian Institute of Technology Mandi, Kamand, Mandi-175075 (H.P.) India
}
\affiliation{$^2$Department of Physical Sciences, Indian Institute of Science Education and Research (IISER) Mohali, Sector 81, S. A. S. Nagar, Manauli, PO 140306, India}

\begin{abstract}
The Planar Hall effect (PHE) in topological materials has been a subject of great interest in recent years. Generally, it is understood to originate from the chiral-anomaly (CA) induced charge pumping between doubly degenerate Weyl nodes. However, the occurrence of PHE in the materials with positive and anisotropic orbital magnetoresistance has raised questions about CA being the sole origin of this effect. Here, we report the PHE, magnetoresistance, and thermal transport properties (Seebeck and Nernst coefficients) on the Ag intercalated PdTe$_2$. We observe positive longitudinal magnetoresistance, the linear field dependence of the amplitude of PHE, and a prolate pattern in the parametric plots. The planar Hall resistivity and anisotropic magnetoresitance fits well with theoretical study of CA being the origin of PHE. So, our observations are consistent with Weyl physics dominating the PHE in PdTe$_2$, Cu$_{0.05}$PdTe$_2$, and Ag$_{0.05}$PdTe$_2$. We further support our data with a theoretical model that reproduces the qualitative experimental features. In addition, we have calculated the Seebeck ($\it{S}$) and Nernst ($\nu$) coefficients for PdTe$_2$ and Cu and Ag intercalated compounds. The estimated values of Fermi energy for the Cu and Ag intercalated compounds are respectively two times and three times larger than that of PdTe$_2$. 
\end{abstract}

\maketitle

\section{Introduction}
Topological semimetals (TSMs) have gained enormous research interest among the condensed matter physicist because of their exotic low-temperature properties \cite{shekhar2015extremely, hooda2018unusual, yang2021anomalous, PhysRevB.103.L201101, hooda2022magnetotransport}. In Dirac semimetals, the conduction and valence band meet each other at one k-point. These four-fold degenerate band crossings are called Dirac points and the Dirac fermions disperse linearly in the momentum directions. These fermions are protected by crystal symmetries like time reversal and inversion symmetry, and the breaking of any such symmetry results in the splitting of Dirac point into a pair of doubly degenerate Weyl points with opposite chirality ($\pm$1) and the corresponding semimetal is called Weyl semimetal. Some TSMs, like Na$_3$Bi, Cd$_3$As$_2$, TaAs, TaP, NaAs, NbP, exhibit linear cone-shaped band dispersion and point like Fermi surface \cite{PhysRevB.85.195320, neupane2014observation, weng2015weyl}. These TSMs obey Lorentz invariance and are called type-I TSMs. Meanwhile, there exists another class of TSMs, like Ta$_3$S$_2$, LaAlGe, TaIrTe$_4$, WTe$_2$, WP$_2$, PdTe$_2$ and VAl$_3$, in which an additional kinetic energy term results in the breaking of Lorentz invariance and hence the Dirac/Weyl cones are tilted strongly along a certain momentum direction \cite{chang2016strongly, xu2017discovery, koepernik2016tairte, soluyanov2015type, PhysRevB.96.165134, PhysRevLett.117.066402, PhysRevB.96.041201, yan2017lorentz, PhysRevB.96.125102, chen2018bulk}. These materials with pocket-like Fermi surface configurations are known as type-II TSMs. 
Breaking of Lorentz invariance in type-II TSMs leads to some remarkable properties \cite{zyuzin2016intrinsic, ferreira2021strain, PhysRevLett.116.236401, PhysRevLett.117.077202}. One such property is the Planar Hall effect (PHE), which was initially proposed as a manifestation of chiral anomaly in TSMs \cite{nandy2017chiral}. PHE emerges in a system when the magnetic field (\textit{H}) is applied in the same plane as that of electric current (\textit{I}) and the induced transverse voltage. Since the conventional Hall effect vanishes in this configuration, the PHE is believed to be a helpful tool to identify and ascertain a system's topological nontrivial band structure. 

The group-X transition metal dichalcogenide PdTe$_2$ has shown some intriguing planar Hall behavior. PdTe$_2$ is a type-II Dirac semimetal and shows superconductivity below $\sim$ 1.7 K \cite{PhysRevB.97.054515, kumar2022investigation}. The PHE reports on PdTe$_2$ by Xu \textit{et al.} \cite{xu2018planar} and Meng \textit{et al.} \cite{meng2019planar} have claimed the origin of PHE to be the chiral anomaly and orbital MR respectively. Earlier, we studied the PHE in 5$\%$ Cu intercalated PdTe$_2$, where we observed the presence of positive orbital MR \cite{sonika2021planar}. However, in another report on Cu intercalated system by Feng \textit{et al.}, the PHE has been attributed to chiral anomaly \cite{feng2022planar}. Such intriguing observations on PdTe$_2$ systems engrossed us for further exploration of PHE in PdTe$_2$ and Ag$_{0.05}$PdTe$_2$ in order to understand the origin of PHE in these systems and to see the effect of Ag intercalation on the transport properties. 

The PHE has been a topic of discussion for the past few years due to its different origins in different systems \cite{taskin2017planar, he2019nonlinear,  PhysRevB.101.041408, liang2018experimental,  kumar2018planar, liu2019nontopological}. Its origin in topological semimetals is of special interest because of its association with chiral anomaly \cite{nandy2017chiral, PhysRevB.107.L081110}. Chiral anomaly was generally thought to be accompanied by a negative longitudinal magnetoresistance (LMR) due to the generation of chiral current between Weyl nodes with opposite chirality when the magnetic field and electric field are parallel to each other and it was believed that the semiclassical theory, which generally conjectures a spherical or an ellipsoidal Fermi surface, cannot explain the PHE originating in systems with positive LMR. Systems showing PHE with positive LMR have posed a question about chiral anomaly being the origin of PHE \cite{liu2019nontopological, meng2019planar, li2020planar}.  

\begin{figure*}
\includegraphics[width= 18 cm, height = 11 cm]{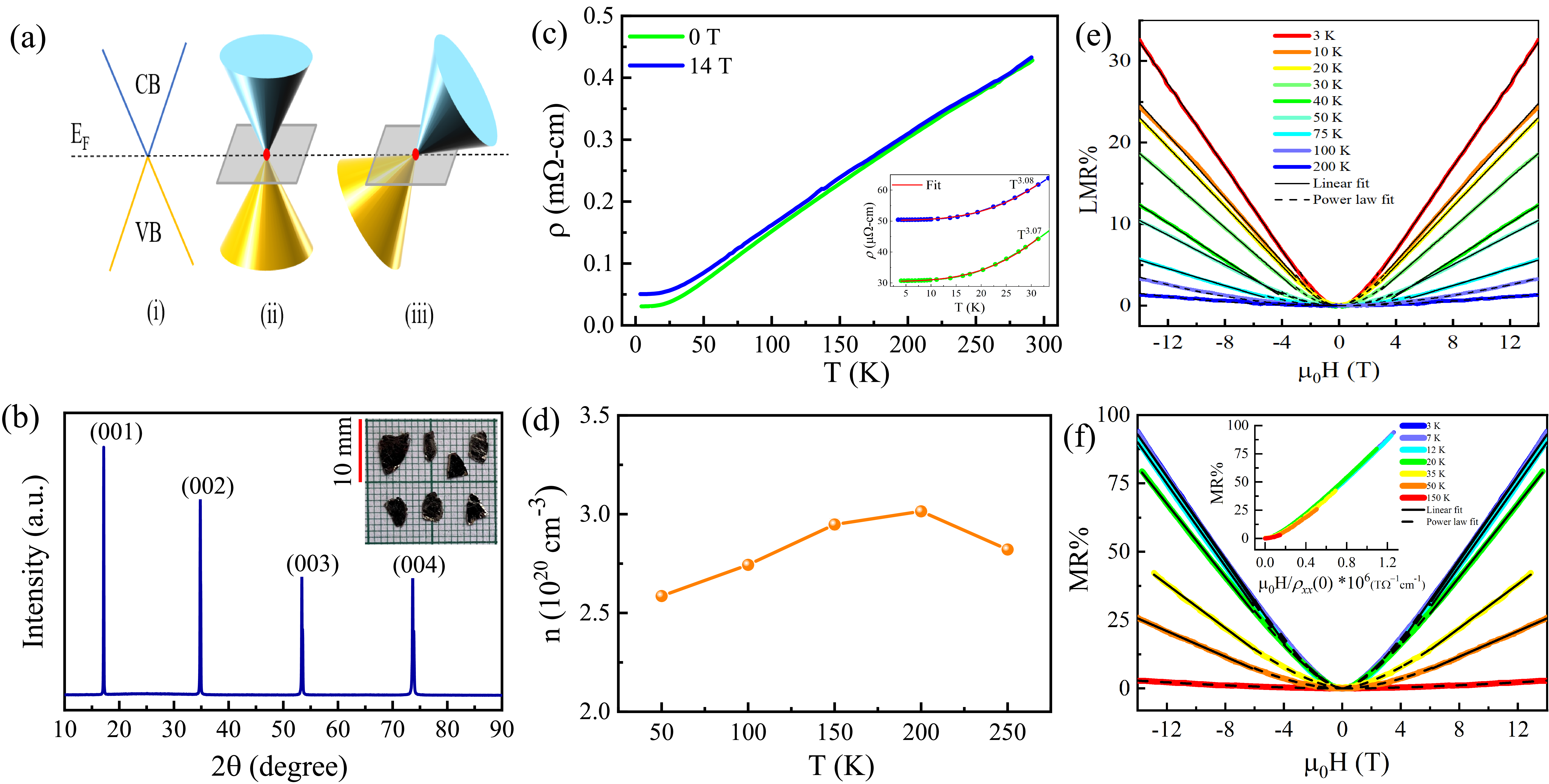}
\caption{(a) Schematic band structure of (i) 3D Dirac semimetal (ii) Type-I and (iii) Type-II Dirac semimetal. E$_F$ denotes the Fermi energy level, and CB and VB are the conduction and valence bands, respectively. The gray color plane represents the Fermi level and the squares show the Fermi arcs. (b) Single crystal X-ray diffraction pattern of Ag$_{0.05}$PdTe$_2$ showing (00l) reflections. Inset shows the single crystalline samples of Ag$_{0.05}$PdTe$_2$. The smallest grid size in the inset is 1 mm. (c) Electrical resistivity at 0 and 14 Tesla field. Inset shows the $\sim{T^3}$ fit for T = 3 - 30 K. (d) Temperature dependence of carrier concentration. (e) MR measured for \textit{T} = 3 - 200 K when $\it{\mu_0 H \parallel I}$. (f) MR measured for \textit{T} = 3 - 150 K when $\it{\mu_0 H} \perp \it{I}$. The inset shows Kohler's plot.}
\label{fig:Figure1}
\end{figure*} 

In the present study, we explore the PHE, anisotropic magnetoresistance (AMR) and thermal transport (Seebeck and Nernst coefficients) properties of Ag$_{0.05}$PdTe$_2$ and PdTe$_2$ single crystals to address the generic nature of these properties in the pristine PdTe$_2$. We show that the magnetoresistance data in Ag$_{0.05}$PdTe$_2$ and PdTe$_2$ fully agrees with the expectations of Weyl physics. We support our studies with a theoretical model that qualitatively explains the observed experimental features. We observe a large non-saturating MR, which is a combination of linear and parabolic field dependence. The amplitude of PHE shows linear magnetic field dependence for 0 - 14 Tesla. A large value of AMR ($\sim$ 50 $\mu\Omega$-cm for \textit{T} = 2.5 K and $\it{\mu_0 H}$ = 14 Tesla) has been obtained for Ag$_{0.05}$PdTe$_2$. The AMR ratio in Ag$_{0.05}$PdTe$_2$ is much larger than those of traditional ferromagnetic metals. The thermoelectric response of Cu and Ag intercalated PdTe$_2$ is compared with the parent compound PdTe$_2$. The intercalation results in enhancing Fermi temperature and the values of Seebeck and Nernst coefficients.

\section{Experimental section}

Single crystalline samples of Ag$_{0.05}$PdTe$_2$ were prepared in two step route. First, the polycrystalline samples were prepared by solid state reaction route, taking the granules of Ag (99.99 $\%$), Pd (99.95 $\%$), and Te (99.999 $\%$) in stoichiometric ratio. The polycrystalline sample was then grounded, pelletized and sealed in evacuated quartz tube with pressure lower than ${10}^{-4}$ mbar. In the second step, the tube containing the sample was heated to ${850} ^o$C for 48 hours and then cooled to ${550} ^o$C at very slow cooling rate of ${2.5} ^o$C per hour. It was kept at ${550} ^o$C for another 48 hours and then naturally cooled down to room temperature. We could obtain a large number of single crystals of dimensions 2 mm x 5 mm (Inset of figure \ref{fig:Figure1} (b)). Figure \ref{fig:Figure1} (b) show the c-axis [\textit{00l}] orientation of the single crystal flakes in the X-ray diffraction (XRD) pattern. 
 
Room temperature XRD measurements of powdered sample and single crystal flake were performed on Rigaku Smartlab X-Ray Diffractometer and Full Prof Rietveld refinement method has been used for analysis. Room temperature XRD pattern confirms hexagonal structure (space group: {\it P}$\bar{3}$m1) for Ag$_{0.05}$PdTe$_2$ (Figure 1 of supplementary material \cite{supp_APT}). The electronic transport measurements were performed using Quantum Design Physical Property Measurement System (PPMS). The angle dependent measurements were performed by rotating the sample in such a way that the in-plane magnetic field makes an angle $\theta$ with the direction of electric current. We have used 1 mA alternating current of the 143.3 Hz frequency for all the measurements. For the Seebeck and Nernst coefficient measurements, we have used a home built experimental set up which is integrated with the PPMS system \cite{sharma2020experimental}. 
 
\section{Results and Discussion}

 \subsection{Magnetoresistance}

 The electrical resistivity ($\rho$(T)) of Ag$_{0.05}$PdTe$_2$ at $\it{\mu_0 H}$ = 0 and 14 Tesla is shown in figure \ref{fig:Figure1}(c). Unlike Cu$_{0.05}$PdTe$_2$, the zero field $\rho$(T) does not show any signature of superconductivity down to \textit{T} = 2 K. Interestingly the $\rho$(T) shows a $\it{T}^3$ dependence for $\it{T}$ = 2 - 30 K. It is to mention that we have observed a rare $\it{T}^4$ dependence for PdTe$_2$ and Cu$_{0.05}$PdTe$_2$ in the \textit{T} range of 2 - 40 K, corresponding to the anomalous charge transport in these compounds \cite{hooda2018unusual}. Similar $\it{T}^3$ dependence of $\rho$(T) is observed in the case of dominant intervalley scatterings in the Arsenic and Antimony system \cite{bansal1973temperature}. The Hall measurement shows dominant electron type carrier density (\textit{n}) of $\sim$ 10$^{20}$ cm$^{-3}$ (Figure \ref{fig:Figure1}(d)). This \textit{n} value is higher than that for PdTe$_2$ (10$^{22}$ cm$^{-3}$) and Cu$_{0.05}$PdTe$_2$ (10$^{22}$ cm$^{-3}$) indicating the charge transfer from intercalated atom to parent compound \cite{hooda2018electronic, sonika2021planar}. 
 
The longitudinal and transverse magnetoresistance (MR = ($\rho$($\it{\mu_0 H}$) - $\rho$(0))/$\rho$(0)) for $\it{\mu_0 H} \parallel \it{I}$ and $\it{\mu_0 H} \perp \it{I}$ are measured for \textit{T} range 3 - 200 K and 3 - 150 K respectively for $\it{\mu_0 H}$ =  0 - 14 Tesla (shown in figure \ref{fig:Figure1}(e), (f)). Both longitudinal and transverse MR are positive for all the measured temperatures and their values increase on lowering \textit{T}. The longitudinal MR shows power law field dependence ($\sim$ a$\it{(\mu_0 H)^m}$; where a is the proportionality constant and \textit{m} = 1.5 - 1.7 is the power index) at high \textit{T} (100 - 200 K) for the complete $\it{\mu_0 H}$ range of 0 - 14 Tesla. For the low \textit{T} (= 3 - 75 K), MR shows non saturating linear $\it{\mu_0 H}$ dependence for $\it{\mu_0 H} >$ 4 Tesla. Similarly for transverse MR, linear $\it{\mu_0 H}$ dependence is observed for $\it{T}$ = 3 - 50 K and $\it{\mu_0 H} >$ 6 Tesla, and $\sim$ a$\it{(\mu_0 H)^m}$; (\textit{m} = 1.4 - 1.6) for $\it{\mu_0 H}$ = 0 - 6 Tesla. Generally, the linear MR is explained by Abrikosov model, which relates the linear MR to the linear energy dispersion of the Dirac fermions and is applicable in the extreme quantum limit, when $\it{\mu_0 H}$ is quite large such that all Landau level are well formed and the carrier concentration is small enough so that electrons occupy only the lowest Landau level\cite{PhysRevB.58.2788}. The transverse MR value of $\sim$ 93 $\%$ is observed at $\it{\mu_0 H}$ = 14 Tesla and \textit{T} = 3 K. Such high values of MR in a non-magnetic system may arise due to complexity of Fermi surface and macroscopic inhomogeneities. Also, anisotropy of Fermi surface is evident from large difference in the values of transverse and longitudinal MR at same value of $\it{T}$ and $\it{\mu_0 H}$. Further we analyzed the transverse MR data by employing Kohler’s scaling at different temperatures. The change in $\rho$ in an applied $\it{\mu_0 H}$ depends on $\omega_C \tau$ where $\omega_C (\varpropto \it{\mu_0 H}$) is the cyclotron frequency and $\tau(T) (\varpropto 1/\rho(T)$) is the relaxation time. Inset of figure \ref{fig:Figure1}(f) shows the Kohler's scaling plots where all the MR curves at different \textit{T} collapse onto a single curve suggesting a single scattering mechanism in the Ag$_{0.05}$PdTe$_2$.

\subsection{Planar Hall effect and Anisotropic magnetoresistance}

\begin{figure}[tb]
\includegraphics[width= \columnwidth, height = 10 cm]{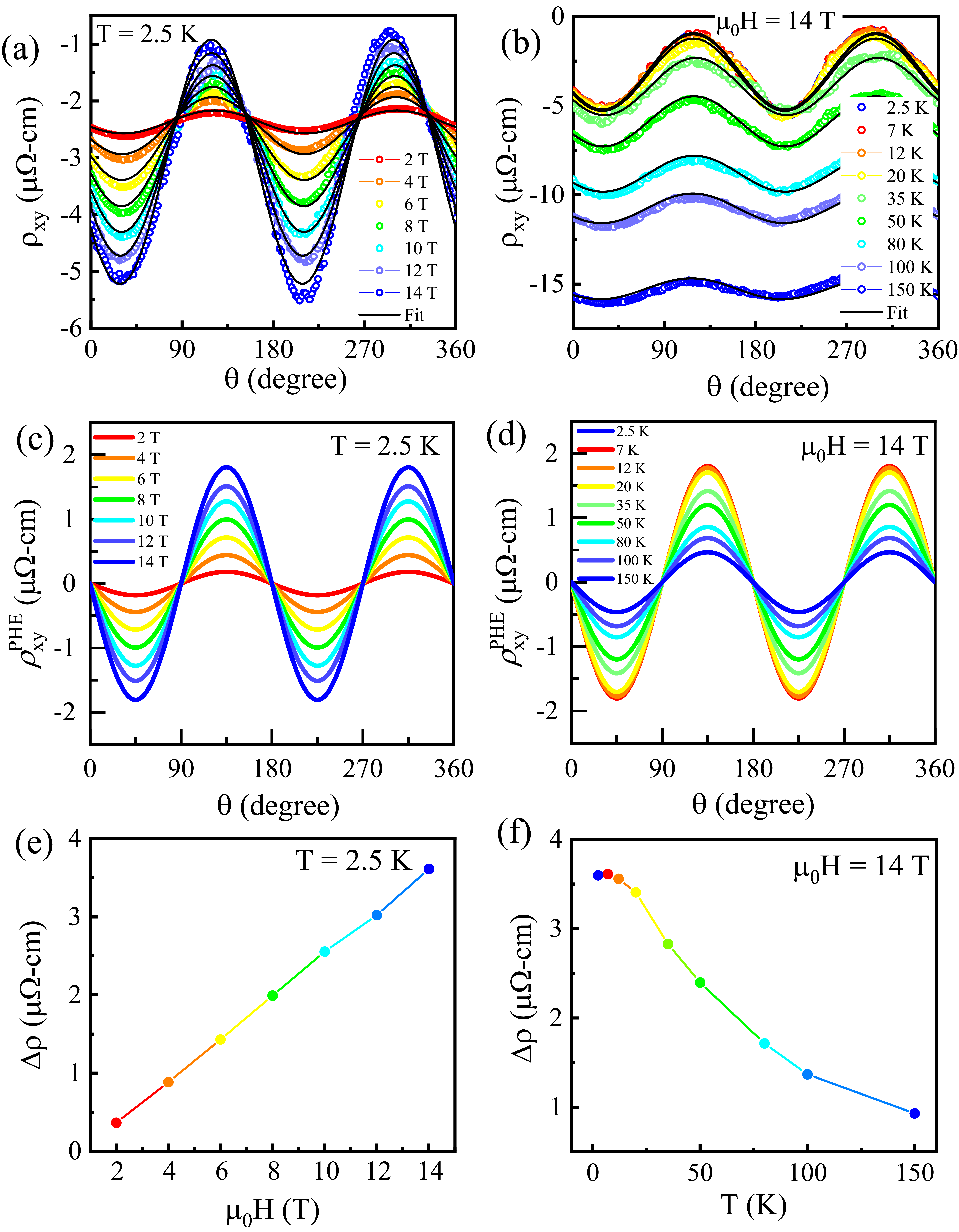}
\caption{(a) Angular dependence of $\rho_{xy}$ measured in planar Hall configuration at different $\it{\mu_0 H}$ for \textit{T} = 2.5 K and (b) at different \textit{T} for $\it{\mu_0 H}$ = 14 Tesla for Ag$_{0.05}$PdTe$_2$. Fitting of data is shown by black curves. Extracted value of planar Hall signal (c) at different $\it{\mu_0 H}$ for \textit{T} = 2.5 K and (d) at different \textit{T} for $\it{\mu_0 H}$ = 14 Tesla. The (e) $\it{\mu_0 H}$ and (f) \textit{T} dependence of PHE amplitude $\Delta\rho$}
\label{fig:Figure2}
\end{figure}

Figure \ref{fig:Figure2}(a) and \ref{fig:Figure2}(b) represent the angular dependence of transverse resistivity ($\rho_{xy}$) at different $\it{\mu_0 H}$ for $\it{T}$ = 2.5 K and at different \textit{T} for $\it{\mu_0 H}$ = 14 Tesla, respectively for Ag$_{0.05}$PdTe$_2$. The observed $\rho_{xy}$ has contribution from three factors; the planar Hall component ($\rho^{PHE}_{xy}$), the longitudinal component due to misalignment of Hall contacts ($\rho_{long.}$) and the constant resistivity component $\rho_{G}$ due to non-uniform thickness of the sample \cite{nandy2017chiral, burkov2017giant}

\begin{equation}
\rho_{xy} = \rho^{PHE}_{xy} + \rho_{long.} + \rho_{G}
\label{eq1}
\end{equation}

where $\rho^{PHE}_{xy} = -\Delta\rho sin\theta cos\theta$, $\rho_{long.} = \Delta\rho cos^2\theta$ and $\Delta\rho$ = $\rho_{\perp}$ - $\rho_{\parallel}$ is the amplitude of PHE. $\rho_{\parallel}$ and $\rho_{\perp}$ are the resistivities corresponding to $\theta$ is $0^o$ and $90^o$ respectively.

 The planar Hall resistivity at different $\it{\mu_0 H}$ for $\it{T}$ = 2.5 K and at different \textit{T} for $\it{\mu_0 H}$ = 14 Tesla is shown in figure \ref{fig:Figure2}(c) and \ref{fig:Figure2}(d) respectively. The magnetic field dependence of $\Delta\rho$ extracted from fitting of $\rho_{xy}$ using equation \ref{eq1} is shown in figure \ref{fig:Figure2}(e). The value of $\Delta\rho$ increases linearly with $\it{\mu_0 H}$. For the chiral anomaly driven PHE, its amplitude $\Delta\rho$, is expected to follow quadratic behavior at low $\it{\mu_0 H}$, followed by the linear behavior at high $\it{\mu_0 H}$, depending on the strength of the length scale related to magnetic field (\textit{$L_a$}) and chiral charge diffusion length (\textit{$L_c$})\cite{burkov2017giant}. In the weak magnetic field limit (\textit{$L_a$}$\gg$\textit{$L_c$} ), the $\Delta\rho$ follows quadratic field dependence as per following relation;
 
\begin{equation}
\rho^{PHE}_{xy} \varpropto \left(\frac{L_c}{L_a}\right)^2 sin\theta cos\theta \varpropto\textit{($\mu_0$H)$^2$} 
\label{eq2}
\end{equation}

where \textit{$L_a$} = D/$\it{\mu_0 H}\Gamma$ ; \textit{$L_c$} = (D$\tau_c)^{1/2}$ and the parameters D, $\Gamma$, and $\tau_c$ are the diffusion coefficient, transport coefficient, and the scattering time of chiral charge respectively. It is to note that the strength of chiral anomaly induced magnetotransport effects is determined by the ratio of two length scales \textit{i.e.} \textit{$L_c$}/\textit{$L_a$}. For strong $\it{\mu_0 H}$, \textit{$L_a$} $\le$ \textit{$L_x$} $\le$ L$_c^2$/L$_a$ (\textit{$L_x$} is the length of the sample), $\Delta\rho$ follows the relation \cite{burkov2017giant},

\begin{equation}
\rho^{PHE}_{xy} \varpropto \left(\frac{L_a}{L_x}\right) sin\theta cos\theta \varpropto\textit{$\mu_0$H}
\label{eq3}
\end{equation}
 
This type of $\Delta\rho$ behavior (quadratic and linear $\it{\mu_0 H}$ dependence in the low and high $\it{\mu_0 H}$ regions respectively) has been observed previously for Te and MoTe$_2$ \cite{zhang2020magnetotransport, chen2018planar}. The factor 1/\textit{$L_a$} determines the strength of trivial and chiral charges so the quadratic and linear field dependence of $\Delta\rho$ shows weak and intermediate coupling of electric and chiral charges respectively as per equation \ref{eq2} and \ref{eq3}. The coupling strength of electric and chiral charges can be further confirmed from the temperature dependence of $\Delta\rho$. There are systems in which Fermi surface reconstruction results in an abrupt upturn in $\Delta\rho (T)$ at certain \textit{T} at which there is anomaly in the $\rho (T)$ accompanied by some other transitions like change in density of states and mass anisotropies \cite{chen2018planar, yan2020giant, yang2021planar}. However, we do not observe any signature of either the coupling of electric and chiral charge or the Fermi surface reconstruction for Ag$_{0.05}$PdTe$_2$ system. Therefore, the linear behavior of $\Delta\rho (\it{\mu_0 H})$ can be attributed to factors like Fermi surface anisotropies (FSA), carrier concentration, or the mass anisotropies.

Generally, in the absence of Lorentz force ($\it{\mu_0 H} \parallel \it{I}$), both $\rho_{\parallel}$ and $\rho_{\perp}$ should be equal to the zero-field resistivity making $\Delta\rho$ (= $\rho_{\perp}$ - $\rho_{\parallel}$) zero. So, ideally, PHE should be zero and AMR should be a constant. However, due to the presence of chiral anomaly in Dirac/Weyl semimetals, $\rho_{\parallel}$ shows a negative $\it{\mu_0 H}$ dependence which results into a non-zero value of $\Delta\rho$ and periodic angular variation of PHE and AMR. So, the PHE component propagates as per equation \ref{eq1} with the minima and maxima at ${45}^o$ and ${135}^o$ respectively with an oscillation period of $\pi$. 
There can be several reasons for PHE, such as anisotropic orbital MR \cite{meng2019planar, liu2019nontopological, li2020planar}, spin momentum locking \cite{sulaev2015electrically, rao2021theory}, anisotropic magnetic scattering \cite{PhysRevB.101.041408, zhang2020influence}, other than the chiral anomaly. Ag$_{0.05}$PdTe$_2$ being a non-magnetic material, directly rule out the possibility of magnetic origin for PHE. Our theoretical studies (Section IV), reveals that at some finite intervalley scattering the longitudinal MR turns to positive value even in the presence of chiral anomaly.

\begin{figure}[tb]
\includegraphics[width= \columnwidth, height = 10 cm]{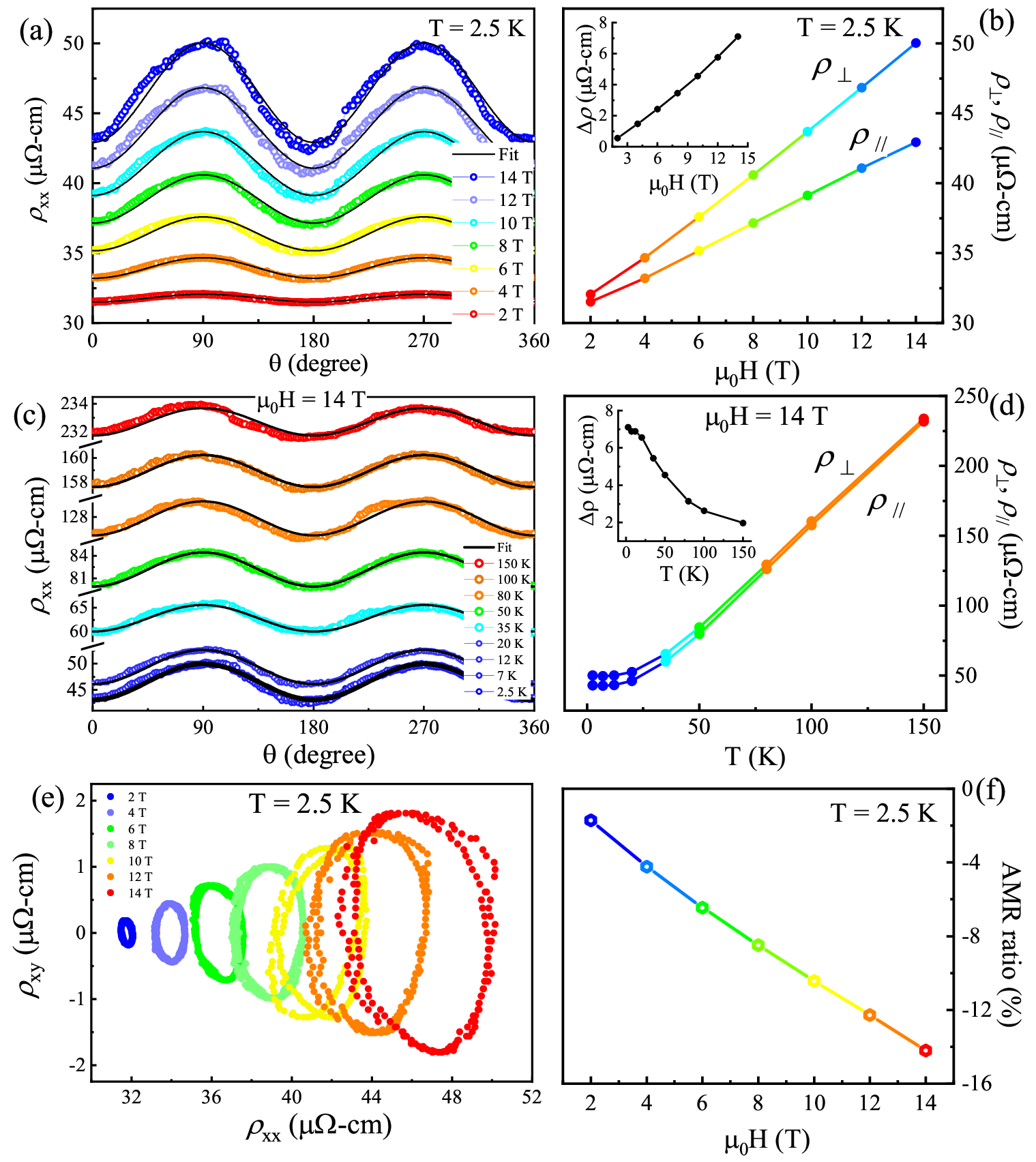}
\caption{Angular dependence of $\rho_{xx}$ (a) at different $\it{\mu_0 H}$ for \textit{T} = 2.5 K and (c) at different \textit{T} for $\it{\mu_0 H}$ = 14 Tesla. The (b) $\it{\mu_0 H}$ and (d) \textit{T} dependence of $\rho_\perp$ and $\rho_\parallel$ extracted from fitting of $\rho_{xx}$. The corresponding insets are for $\Delta\rho$. (e) Parametric plots of the planar Hall and AMR signals showing the orbits at \textit{T} = 2.5 K. (f) The amplitude of AMR ratio at different magnetic fields for \textit{T} = 2.5 K.}
\label{fig:Figure3}
\end{figure}
 
Figure \ref{fig:Figure3}(a) and \ref{fig:Figure3}(c) shows AMR at different $\it{\mu_0 H}$ for $\it{T}$ = 2.5 K and at different \textit{T} for $\it{\mu_0 H}$ = 14 Tesla respectively for Ag$_{0.05}$PdTe$_2$. The AMR data is fitted using following equation \cite{nandy2017chiral, burkov2017giant} 

\begin{equation}
\rho_{xx} = \rho_{\perp} - \Delta\rho cos^2\theta
\label{eq4}
\end{equation}
 
The $\it{\mu_0 H}$ and \textit{T} dependence of $\rho_{\parallel}$ and $\rho_{\perp}$ obtained from the fitting of $\rho_{xx}$ are shown in figure \ref{fig:Figure3}(b) and \ref{fig:Figure3}(d) respectively. With the increase in $\it{\mu_0 H}$, both $\rho_{\parallel}$ and $\rho_{\perp}$ increases causing $\Delta\rho$ to increase with the increase in $\it{\mu_0 H}$ (inset of figure \ref{fig:Figure3}(b)). Theoretically, $\rho_{\perp}$ is assumed to be independent of $\it{\mu_0 H}$ and $\rho_{\parallel}$ should be decreasing with the increase in $\it{\mu_0 H}$ as is observed for WTe$_2$ and ZrSiSe \cite{meng2019planar, bo2019observation}. The observed $\it{\mu_0 H}$ dependency of $\rho_{\perp}$ and $\rho_{\parallel}$ for Ag$_{0.05}$PdTe$_2$ suggest the presence of orbital MR in the system due to anisotropy of transport parameters such as effective mass, scattering time, mobility etc., which arises from the FSA and vary along different directions \cite{meng2019planar, liu2019nontopological, roy2021signature}. The \textit{T} dependence of $\Delta\rho$ is shown in the inset of figure \ref{fig:Figure3}(d) which is similar to that obtained from planar Hall resistivity. 

In order to confirm the chiral anomaly, we plotted the values of $\rho^{PHE}_{xy}$ against the values of anisotropic resistivity $\rho_{xx}$ at different $\it{\mu_0 H}$ keeping $\theta$ as a parameter (figure \ref{fig:Figure3}(e)). This plot is called the parametric plot. In these parametric plots, the orbits are expanding in the form of prolates towards large $\rho_{xx}$ without showing any saturation up to $\it{\mu_0 H}$ = 14 Tesla. Some compounds like Na$_3$Bi, GdPtBi etc. show concentric circles and some other compounds like PtSe$_2$ and PtTe$_2$ show shock wave pattern in parametric plot \cite{liang2018experimental, li2020planar, liu2019nontopological}. The prolate pattern obtained for Ag$_{0.05}$PdTe$_2$ conform with that generated through a theoretical model (section IV) asssuming chiral anomaly causing PHE.

\begin{figure}[tb]
\includegraphics[width= 7 cm, height = 5 cm]{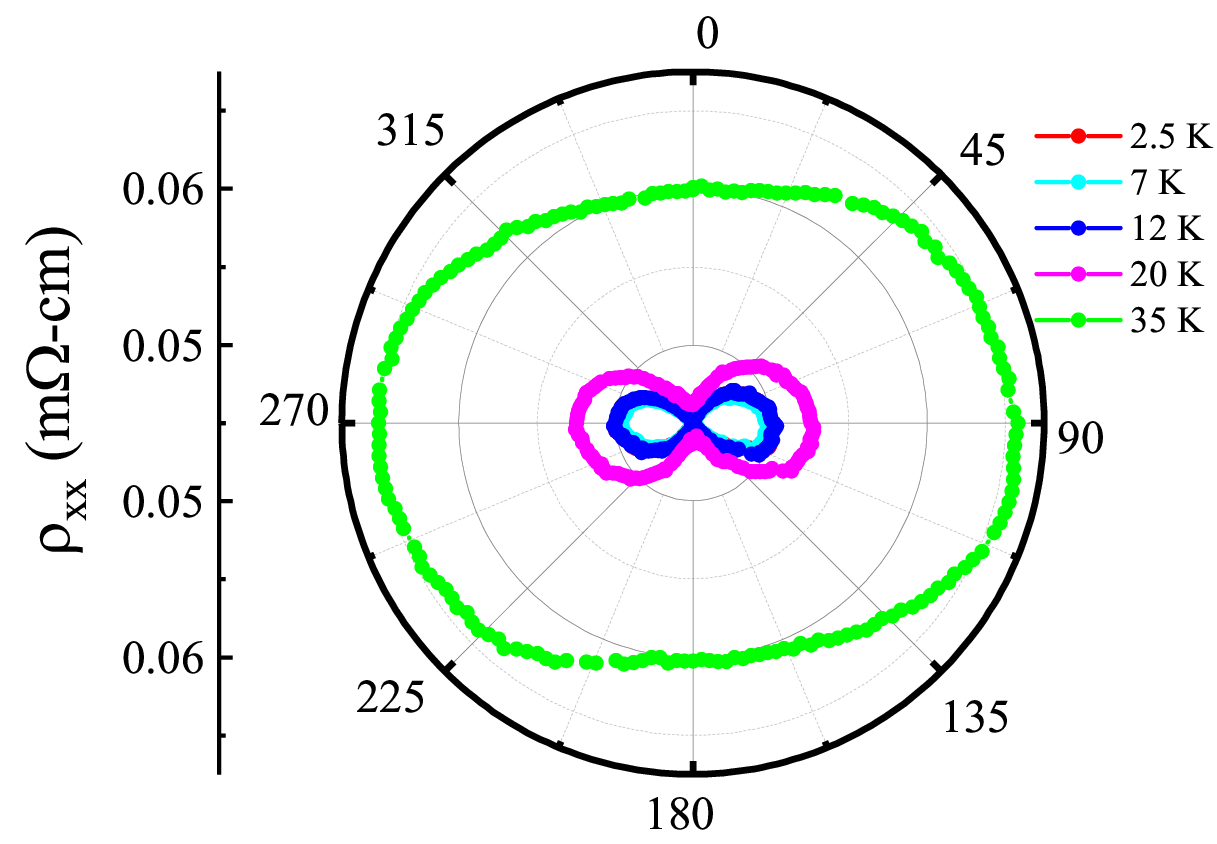}
\caption{Polar plot of angular dependence of $\rho_{xx}$ taken at 14 Tesla for temperatures 2.5 - 35 K}
\label{fig:Figure4}
\end{figure}

\begin{figure}[tb]
\includegraphics[width= \columnwidth, height = 9 cm]{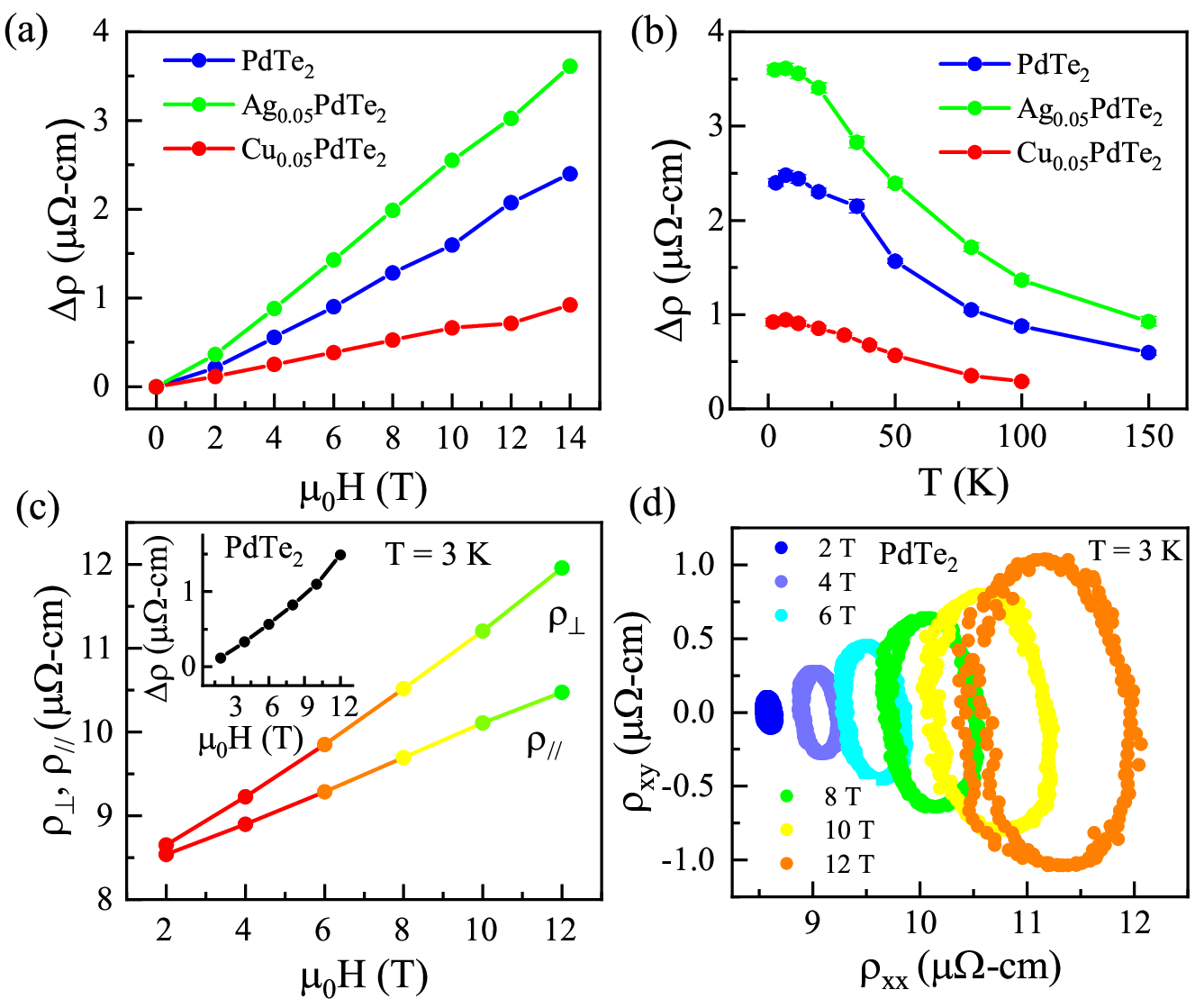}
\caption{The (a) $\it{\mu_0 H}$ and (b) \textit{T} dependence of amplitude of PHE $\Delta\rho$ for PdTe$_2$, Cu$_{0.05}$PdTe$_2$, and Ag$_{0.05}$PdTe$_2$ (c) The $\it{\mu_0 H}$ dependence of $\rho_\perp$ and $\rho_\parallel$ extracted from fitting of $\rho_{xx}$ for PdTe$_2$ (d) Parametric plot of the planar Hall and AMR signals showing the orbits at \textit{T} = 3 K for PdTe$_2$.}
\label{fig:Figure5}
\end{figure}

Figure \ref{fig:Figure3}(f) shows the magnetic field dependence of AMR ratio (AMR$\%$ = ($\rho_{\parallel}$ - $\rho_{\perp}$)/$\rho_{\perp}$ $\times$ 100$\%$) for Ag$_{0.05}$PdTe$_2$. The magnitude of AMR ratio is found to increase with the increase in $\it{\mu_0 H}$. A large AMR ratio of $\sim$ -14$\%$ has been obtained for \textit{T} = 2.5 K and $\it{\mu_0 H}$ = 14 Tesla. This value of AMR ratio in Ag$_{0.05}$PdTe$_2$ is much larger in comparison to that of many ferromagnetic metals \cite{kokado2012anisotropic, sakuraba2014quantitative, sato2018large}, and  makes it a promising candidate in building magnetic sensors and magnetoresistive devices with low power consumption.

Figure \ref{fig:Figure4} shows the polar plot of angular dependence of AMR measured at constant magnetic field of 14 Tesla for different temperatures (\textit{T} = 2.5, 7, 12, 20, 35 K). We observed that the AMR value is maximum when $\it{\mu_0 H}\perp\textit{I}$ and becomes minimum when $\it{\mu_0 H}\parallel\textit{I}$. The angular dependence of $\rho_{xx}$ shows two-fold anisotropy up to 20 K and above 20 K it does not show any difference for different angles. These observations are consitent with the temperature dependence of $\Delta\rho$ shown in figure \ref{fig:Figure2}(f) pointing towards the same mechanism for AMR and PHE in this system. 

In the figure \ref{fig:Figure5}(a) and \ref{fig:Figure5}(b), we have plotted the $\it{\mu_0 H}$ and \textit{T} dependence of amplitude of PHE respectively for our PdTe$_2$ and Ag$_{0.05}$PdTe$_2$, with the values for Cu$_{0.05}$PdTe$_2$ from literature \cite{sonika2021planar}. The values of $\Delta\rho$ for our PdTe$_2$ and Ag$_{0.05}$PdTe$_2$ are comparble to that of PdTe$_2$ reported by Meng \textit{et al.} \cite{meng2019planar} and are one order of magnitude larger than those reported by Xu \textit{et al.} \cite{xu2018planar} and Sonika \textit{et al.} \cite{sonika2021planar}. The $\Delta\rho$ values for Cu$_{0.05}$PdTe$_2$ reported by Feng \textit{et al.} \cite{feng2022planar} are almost $\sim$ 2 times that of our PdTe$_2$ and Ag$_{0.05}$PdTe$_2$ (Table \ref{Table1}). The AMR values for Ag$_{0.05}$PdTe$_2$ are almost 50 times larger than those for PdTe$_2$ reported by Xu \textit{et al.} and for Cu$_{0.05}$PdTe$_2$ reported by Sonika \textit{et al.}. The magnitude of AMR in Ag$_{0.05}$PdTe$_2$ is almost double the magnitude of AMR for Cu$_{0.05}$PdTe$_2$ reported by Feng \textit{et al.} and is nearly one-fourth of that of Meng \textit{et al.} for PdTe$_2$. The AMR value is $\sim$ 14$\%$ at 2.5 K and 14 Tesla, which is almost six times smaller than transverse MR value of $\sim$ 93$\%$ obtained at same temperature and magnetic field. The chiral anomaly contribution is concealed by orbital MR in such cases \cite{liu2019nontopological, li2020planar, meng2019planar}. The FSA and the existence of Dirac point below Fermi level complicate the separation of contribution of conventional and Dirac fermions. Figure \ref{fig:Figure5}(c) shows the $\it{\mu_0 H}$ dependence of $\rho_{\parallel}$ and $\rho_{\perp}$ extracted from $\rho_{xx}$ for PdTe$_2$ at T = 3 K. The corresponding inset shows the $\it{\mu_0 H}$ dependence of $\Delta\rho$. The parametric plot (figure \ref{fig:Figure5}(d)) and the $\it{\mu_0 H}$ dependence of $\rho_{\parallel}$, $\rho_{\perp}$ and $\Delta\rho$ for PdTe$_2$ is similar to that observed for Ag$_{0.05}$PdTe$_2$ showing common origin of PHE in the parent and intercalated systems. The angular dependence of planar Hall and anisotropic longitudinal resistivity of PdTe$_2$ and Cu$_{0.05}$PdTe$_2$ is shown in the supplementary material \cite{supp_APT}. 

\begin{table}[htb]
\centering
\caption{\label{Table1} Amplitude of planar Hall resistivity and anisotropic longitudinal resistivity at $\it{\mu_0 H}$ = 8 Tesla.}
\begin{tabular}{cccc}
\hline
\textrm{Compound}&
\textrm{$\Delta\rho$ ($\mu\Omega$-cm)}&
\textrm{AMR ($\mu\Omega$-cm)}&
\textrm{Reference}\\
\hline
PdTe$_2$  &  0.067 & 0.75 & \cite{xu2018planar}\\

--  &  1.35   &  142.5  & \cite{meng2019planar}\\

--  &  1.28   &  11.21  &  Present work\\
			
Cu$_{0.05}$PdTe$_2$  &  0.12  & 0.94  & \cite{sonika2021planar} \\

--  &  4 & 21  & \cite{feng2022planar} \\

--  &  0.53  & 6.01  &  Present work\\
			
Ag$_{0.05}$PdTe$_2$  & 1.99 &  40.59  & Present work \\
		
\hline
\end{tabular}
\end{table}

\subsection{Thermal transport}

\begin{figure}[tb]
\includegraphics[width = 8 cm, height = 10 cm]{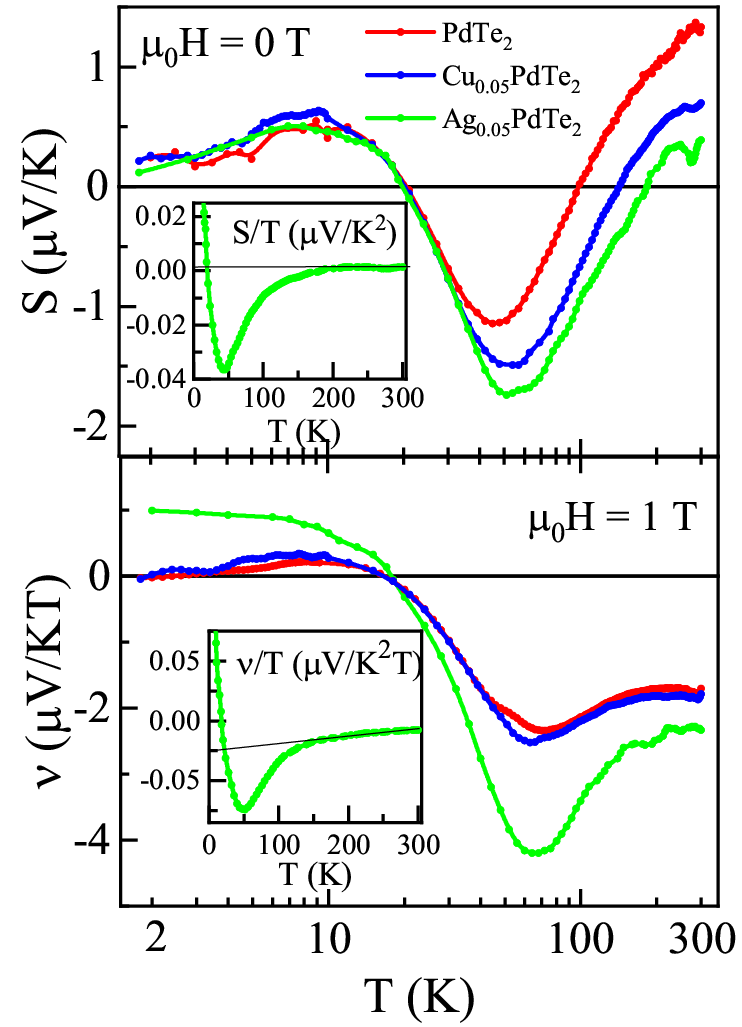}
\caption{Temperature dependence of $\it{S}$ (upper panel) and  $\nu$ (lower panel) of PdTe$_2$, Cu$_{0.05}$PdTe$_2$ and Ag$_{0.05}$PdTe$_2$ at $\it{\mu_0 H}$ = 0 and 1 Tesla respectively. The corresponding insets show the temperature dependence of $\it{S/T}$ and $\nu/T$ respectively.}
\label{fig:Figure6}
\end{figure}

The $\it{T}$ dependence of Seebeck coefficient ($\it{S}$ = $E_x/|\nabla T_x|$) for PdTe$_2$, Cu$_{0.05}$PdTe$_2$, and Ag$_{0.05}$PdTe$_2$ is shown in the upper panel of figure \ref{fig:Figure6}. The room temperature values of $\it{S}$ for PdTe$_2$, Cu$_{0.05}$PdTe$_2$, and Ag$_{0.05}$PdTe$_2$ are $\sim$ 1.33 $\mu$V/K, $\sim$ 0.69 $\mu$V/K and $\sim$ 0.39 $\mu$V/K with a positive to negative crossover temperature of 99 K, 144 K, and 183 K respectively. Generally, the $\it{S(T)}$ in non-magnetic metals is governed by the electron diffusion contribution ($\varpropto T$, dominant at high temperature) and the phonon drag contribution ($\varpropto T^3$, dominant at low temperature). Under the free electron gas approximation, and in the absence of phonon drag, $\it{S(T)}$ shows linear $\it{T}$ dependence as per Mott relation given by \cite{behnia2004thermoelectricity},

\begin{equation}\\
S/T = \pm \frac{\pi^{2}}{2} \frac{k_B}{e} \frac{1}{T_F} = \pm \frac{\pi^{2}}{3} \frac{{k_B}^2}{e} \frac{N(E_F)}{n} 
\label{eq5}
\end{equation}

where $k_B$ is Boltzmann’s constant, e is electronic charge, n is the carrier density, and $T_F$ is the Fermi temperature. The density of states $N (E_F)$ is related to the Fermi energy as $N (E_F) = 3n/2 k_B T_F $. The inset in the upper panel shows the $\it{T}$ dependence of $\it{S/T}$ for Ag$_{0.05}$PdTe$_2$. The zero-temperature extrapolated value of $\it{S/T}$ for PdTe$_2$, Cu$_{0.05}$PdTe$_2$, and Ag$_{0.05}$PdTe$_2$ corresponds to $T_F$ values of 8.3 $\times$ 10$^4$ K, 16.4 $\times$ 10$^4$ K and 30.4 $\times$ 10$^4$ K respectively. The corresponding Fermi energies $E_F$ are 7.19 eV, 14.1 eV and 26.19 eV respectively. The phonon drag minima for $\it{S}$ of three compounds PdTe$_2$, Cu$_{0.05}$PdTe$_2$, and Ag$_{0.05}$PdTe$_2$ occurs at 45 K, 54 K and 51 K respectively. Under free electron gas approximation, the carrier concentration (n) can be estimated using the following relation 

\begin{equation}
E_F = \frac{{(hc)}^{2}}{8 m_e c^2} \left({\frac{3}{\pi}}\right)^{2/3} n^{2/3}
\label{eq6}
\end{equation}

where h is Planck's constant, c is velocity of light, and $m_e$ is mass of electron. The estimated values of n for PdTe$_2$, Cu$_{0.05}$PdTe$_2$, and Ag$_{0.05}$PdTe$_2$ are 8.75 $\times$ 10$^{22}$ cm$^{-3}$, 2.41 $\times$ 10$^{23}$ cm$^{-3}$ and 6.09 $\times$ 10$^{23}$ cm$^{-3}$ respectively which are in close agreement with literature \cite{hooda2018electronic, sonika2021planar}.

The lower panel of figure \ref{fig:Figure6} shows $\it{T}$ dependence of Nernst coefficient ($\nu$ = N/$\it{\mu_0 H}$, N = $E_y/|\nabla T_x|$) in the presence of magnetic field $\it{\mu_0 H}$ = 1 Tesla for PdTe$_2$, Cu$_{0.05}$PdTe$_2$, and Ag$_{0.05}$PdTe$_2$. The room temperature values of $\nu$ for PdTe$_2$, Cu$_{0.05}$PdTe$_2$, and Ag$_{0.05}$PdTe$_2$ are $\sim$ -1.71 $\mu$V/K-Tesla, $\sim$ -1.79 $\mu$V/K-Tesla and $\sim$ -2.34 $\mu$V/K-Tesla respectively. It is observed that the values of $\nu$ does not change much from PdTe$_2$ to Cu$_{0.05}$PdTe$_2$. However, a distinct change in $\nu$ values is observed on Ag intercalation. The phonon drag minima for Nernst coefficient of three compounds PdTe$_2$, Cu$_{0.05}$PdTe$_2$, and Ag$_{0.05}$PdTe$_2$ occurs at 70 K, 63 K and 68 K respectively. The difference in phonon drag minima in \textit{S} and $\nu$ suggest that there is no contribution of longitudinal component in the Nernst values. The magnitude of both $\it{S}$ and $\nu$ at phonon drag minima increases in going from PdTe$_2$ to Cu$_{0.05}$PdTe$_2$ to Ag$_{0.05}$PdTe$_2$. In addition, a new peak is observed at $\sim{9}$ K in both $\it{S}$ and $\nu$ for the three compounds. Similar peak is observed for pure Bismuth for samples of different sizes at $\sim$ 3 K, which is attributed to phonon drag peak \cite{issi1979phonon}. Within the Boltzmann limit, in a single band picture, following expression relates the Nernst coefficient ($\nu$) with the Fermi temperature $\it{T_F}$ \cite{behnia2009nernst},

\begin{equation}\\
\nu = \frac{N}{\it{\mu_0 H}} = \frac{\pi^{2}}{3} \frac{k_B}{e} \frac{T}{T_F} \mu
\label{eq7}
\end{equation}

where $k_B$ is Boltzmann’s constant, e is electronic charge and $\mu$ is carrier mobility. This formula relates the Nernst effect with the charge carrier mobility. Thus, Ag$_{0.05}$PdTe$_2$ has more Nernst signal value with the high carrier mobility ($\sim$ 5486.74 $cm^2 V^{-1} s^{-1}$ at 5 K) than Cu$_{0.05}$PdTe$_2$ ($\sim$ 187.82 $cm^2 V^{-1} s^{-1}$ at 5 K) and PdTe$_2$ ($\sim$ 64.50 $cm^2 V^{-1} s^{-1}$ at 5 K) with lower carrier mobilty values. The value of $\nu/T$ for PdTe$_2$, Cu$_{0.05}$PdTe$_2$, and Ag$_{0.05}$PdTe$_2$ in zero-temperature limit is $\sim$ -0.0149 $\mu V/K^2$Tesla, $\sim$ -0.0169 $\mu V/K^2$Tesla and $\sim$ -0.0248 $\mu V/K^2$Tesla respectively, which is consistent with the mobility values.
 
\section{Theoretical model}
In this section, we show that the experimental observations are fully consistent with the theoretical expectations of Weyl physics. The theoretical model has been discussed in recent works~\cite{sharma2020sign,ahmad2021longitudinal,ahmad2022longitudinal,sharma2023decoupling}, and we summarize it here for completeness. 
We consider a simple model of a two-node Weyl semimetal given by 
\begin{eqnarray}
    \it{\mu_0 H} = \sum_\eta \sum_\mathbf{k} {\eta\hbar v_F (\mathbf{k}\cdot{\tau})}
    \label{Eq:HWeyl2}
\end{eqnarray}
In the above model, $v$ is the Fermi velocity, ${\tau}$ is the vector of Pauli matrices, and $\eta$ is the chirality of the nodes ($\eta=+1$ or $\eta=-1$). We will employ quasiclassical Boltzmann formalism valid in the limits of perturbative fields. Without loss of generality, we assume that we have an electron-like Fermi surface. The Boltzmann equation for the distribution function $f^\eta_\mathbf{k}$ is
\begin{eqnarray}
\left(\frac{\partial}{\partial t} + \dot{\mathbf{r}}^\eta\cdot \nabla_\mathbf{r}+\dot{\mathbf{k}}^\eta\cdot \nabla_\mathbf{k}\right)f^\eta\mathbf{k} = \mathcal{I}_{\mathrm{coll}}[f^\eta_\mathbf{k}],
\label{Eq_boltz1}
\end{eqnarray}
where $\mathcal{I}_{\mathrm{coll}}$ incorporates the effects of scattering.
In the presence of electric (\textbf{E}) and magnetic fields (\textbf{B}), the dynamics equations for electrons are:
\begin{eqnarray}
\dot{\mathbf{r}}^\eta &= \mathcal{D}^\eta \left( \frac{e}{\hbar}(\mathbf{E}\times \boldsymbol{\Omega}^\chi + \frac{e}{\hbar}(\mathbf{v}^\eta\cdot \boldsymbol{\Omega}^\chi) \mathbf{B} + \mathbf{v}_\mathbf{k}^\eta)\right) \nonumber\\
\dot{\mathbf{p}}^\eta &= -e \mathcal{D}^\chi \left( \mathbf{E} + \mathbf{v}_\mathbf{k}^\eta \times \mu_0\mathbf{H} + \frac{e}{\hbar} (\mathbf{E}\cdot\mathbf{B}) \boldsymbol{\Omega}^\eta \right),
\end{eqnarray}
where $\mathbf{v}_\mathbf{k}^\eta$ is the band-velocity, $\boldsymbol{\Omega}^\eta = -\eta \mathbf{k} /2k^3$ is the Berry curvature, and $\mathcal{D}^\eta = (1+e\mathbf{B}\cdot\boldsymbol{\Omega}^\eta/\hbar)^{-1}$, $\mathbf{m}^\eta_\mathbf{k}$ is the anomalous orbital magnetic moment (OMM). In the presence of an external magnetic field, the energy dispersion is given by $\epsilon^{\eta}_{\mathbf{k}}\rightarrow \epsilon^{\eta}_{\mathbf{k}} - \mathbf{m}^\eta_\mathbf{k}\cdot \mathbf{B}$. 

The collision integral $\mathcal{I}_{\mathrm{coll}}[f^\eta_\mathbf{k}]$ is
\begin{eqnarray}
\mathcal{I}_{\mathrm{coll}}[f^\chi_\mathbf{k}] = \sum_{\chi'}\sum_{\mathbf{k}'} W^{\chi\chi'}_{\mathbf{k},\mathbf{k}'} (f^{\chi'}_{\mathbf{k}'} - f^\chi_\mathbf{k}),
\end{eqnarray}
where the scattering rate $W^{\chi\chi'}_{\mathbf{k},\mathbf{k}'}$ is evaluated by the Fermi's golden rule. Note that we have accounted for both internode and intranode scattering. 

\begin{figure*}
    \centering
    \includegraphics[width=1.99\columnwidth]{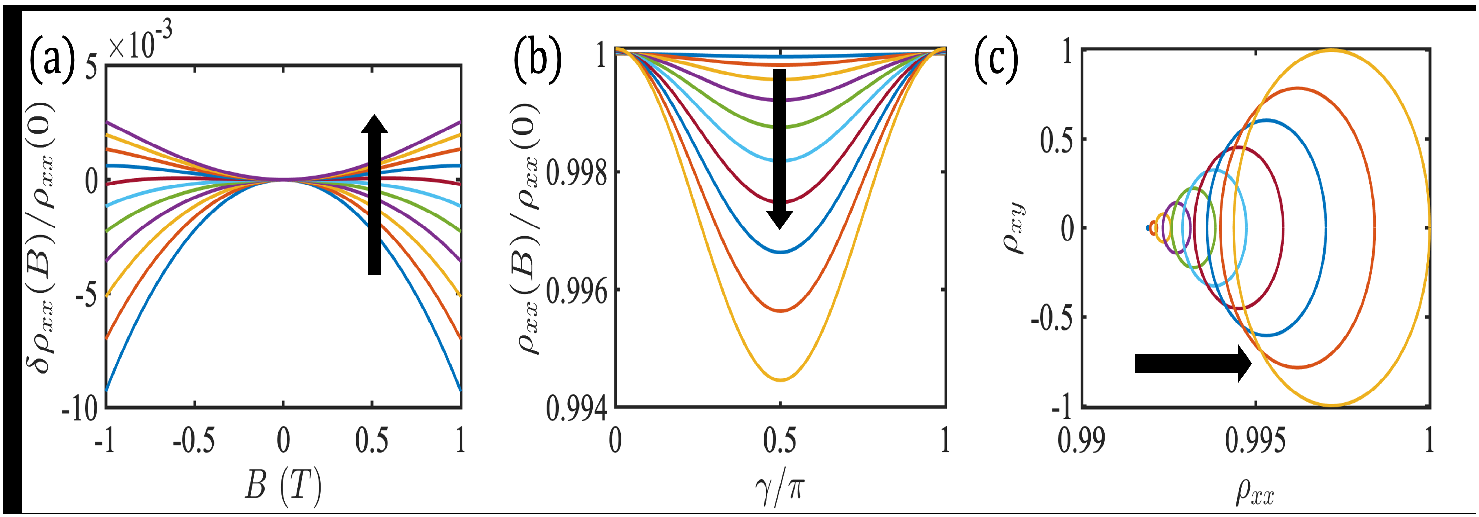}
    \caption{Predictions of magnetoresistivity and Hall resistivity from the theoretical model discussed in Sec. IV. (a) The deviation $\delta \rho_{xx}(B)$ as a function of magnetic field. As we move away from the blue curve in the direction of the arrow, we increase the intervalley scattering strength $\alpha$. Beyond a critical intervalley scattering strength $\alpha_c$, the magnetoresistivity switches sign. (b) Anisotropic magnetoresistance as a function of $\gamma$. The angle $\gamma$ rotates the magnetic field in the $xy-$plane and when $\gamma=\pi/2$, the electric and magnetic field are parallel to each other. (c) Parametric plots of $\rho_{xx}$ vs. $\rho_{xy}$ parametrized by the angle $\gamma$. In (b) and (c) as we move along the direction of arrow, we increase the value of magnetic field from 0.1T to 1T. Plots (b) and (c) are appropriately normalized. }
    \label{fig:theoryplot1}
\end{figure*}

The distribution function takes the form $f^\eta_\mathbf{k} = f_0^\eta + g^\eta_\mathbf{k}$, where $f_0^\chi$ is the equilibrium Fermi-Dirac distribution function and $g^\eta_\mathbf{k}$ is the deviation from equilibrium. 
The steady-state Boltzmann equation (Eq.~\ref{Eq_boltz1}) becomes
\begin{align}
&\left[\left(\frac{\partial f_0^\eta}{\partial \epsilon^\eta_\mathbf{k}}\right) \mathbf{E}\cdot \left(\mathbf{v}^\eta_\mathbf{k} + \frac{e\mathbf{B}}{\hbar} (\boldsymbol{\Omega}^\eta\cdot \mathbf{v}^\eta_\mathbf{k}) \right)\right]\nonumber\\
&= -\frac{1}{e \mathcal{D}^\eta}\sum\limits_{\eta'}\sum\limits_{\mathbf{k}'} W^{\eta\eta'}_{\mathbf{k}\mathbf{k}'} (g^\eta_{\mathbf{k}'} - g^\eta_\mathbf{k})
 \label{Eq_boltz2}
\end{align}
The deviation $g^\eta_\mathbf{k}$ is assumed to be linearly proportional to the electric field, i.e., 
\begin{eqnarray}
g^\eta_\mathbf{k} = e \left(-\frac{\partial f_0^\eta}{\partial \epsilon^\eta_\mathbf{k}}\right) \mathbf{E}\cdot \boldsymbol{\Lambda}^\eta_\mathbf{k}
\end{eqnarray}
We fix the direction of the applied external electric field to be $+\hat{x}$. Therefore, only ${\Lambda}^{\eta x}_\mathbf{k}\equiv {\Lambda}^{\eta}_\mathbf{k}$, is nonzero. Further, the magnetic field is rotated in the $xy-$plane.

Eq.~\ref{Eq_boltz2} up to linear order in $\mathbf{E}$ is:
\begin{eqnarray}
\mathcal{D}^\chi \left[v^{\chi x}_{\mathbf{k}} + \frac{e B}{\hbar}  (\boldsymbol{\Omega}^\chi\cdot \mathbf{v}^\chi_\mathbf{k})\right] = \sum\limits_{\eta}\sum\limits_{\mathbf{k}'} W^{\eta\chi}_{\mathbf{k}\mathbf{k}'} (\Lambda^{\chi}_{\mathbf{k}'} - \Lambda^\eta_\mathbf{k})
\label{Eq_boltz3}
\end{eqnarray} 
We now define 
\begin{eqnarray}
\frac{1}{\tau^\chi_\mathbf{k}} = \mathcal{V} \sum\limits_{\eta} \int{\frac{d^3 \mathbf{k}'}{(2\pi)^3} (\mathcal{D}^\eta_{\mathbf{k}'})^{-1} W^{\eta\chi}_{\mathbf{k}\mathbf{k}'}}
\label{Eq_tau11}
\end{eqnarray}
Substituting the scattering rate in the above equation, we have 
\begin{align}
\frac{1}{\tau^\chi_\mathbf{k}} &= \frac{\mathcal{V}N}{8\pi^2 \hbar} \sum\limits_{\eta} (U^{\chi\eta})^2 \iiint q^2 \sin \theta' \mathcal{G}^{\chi\eta}(\theta,\phi,\theta',\phi') \nonumber\\
&\delta(\epsilon^{\eta}_{\mathbf{q}}-\epsilon_F)(\mathcal{D}^\eta_{\mathbf{q}})^{-1}dq d\theta'd\phi' ,\nonumber\\
\label{Eq_tau1}
\end{align}
where $N$ is the total number of impurities, and $ \mathcal{G}^{\chi\eta}(\theta,\phi,\theta',\phi') = (1+\chi\eta(\cos\theta \cos\theta' + \sin\theta\sin\theta' \cos(\phi-\phi')))$. The Fermi contour $k^\eta$ is evaluated by equating the energy expression to the Fermi energy, and thus 
the three-dimensional integral is  reduced to just integration over the radial variables. The scattering time ${\tau^\chi_\mathbf{k}}$ depends on the chemical potential ($\mu$), and the angular variables  

\begin{align}
&\frac{1}{\tau^\chi_\mu(\theta,\phi)} \nonumber\\
&=\mathcal{V} \sum\limits_{\eta} \iint{\frac{\beta^{\chi\eta}(k')^3}{\mathrm{abs}(\mathbf{v}^\eta_{\mathbf{k}'}\cdot \mathbf{k}'^\eta)}\sin\theta'\mathcal{G}^{\chi\eta}(\mathcal{D}^\eta_{\mathbf{k}'})^{-1} d\theta'd\phi'},
\label{Eq_tau2}
\end{align}
where $abs$ indicates the absolute value.
The Boltzmann equation  now becomes 
\begin{align}
&h^\chi_\mu(\theta,\phi) + \frac{\Lambda^\chi_\mu(\theta,\phi)}{\tau^\chi_\mu(\theta,\phi)} =
\nonumber\\
&\mathcal{V}\sum_\eta \iint {\frac{\beta^{\chi\eta}(k')^3}{\mathrm{abs}(\mathbf{v}^\eta_{\mathbf{k}'}\cdot \mathbf{k}'^\eta)} \sin\theta'\mathcal{G}^{\chi\eta}(\mathcal{D}^\eta_{\mathbf{k}'})^{-1}\Lambda^\eta_{\mu}(\theta',\phi') d\theta'd\phi'}.\nonumber\\
\label{Eq_boltz4}
\end{align}
 We assume the following ansatz for $\Lambda^\chi_\mu(\theta,\phi)$
\begin{align}
&\frac{\Lambda^\chi_\mu(\theta,\phi)}{\tau^\chi_\mu(\theta,\phi)} = \nonumber\\
&(\lambda^\chi - h^\chi_\mu(\theta,\phi) + a^\chi \cos\theta +b^\chi \sin\theta\cos\phi + c^\chi \sin\theta\sin\phi),
\label{Eq_Lambda_1}
\end{align}
where we solve for the eight unknowns ($\lambda^{\pm 1}, a^{\pm 1}, b^{\pm 1}, c^{\pm 1}$). The L.H.S in Eq.~\ref{Eq_boltz4} simplifies to $\lambda^\chi + a^\chi \cos\theta + b^\chi \sin\theta\cos\phi + c^\chi \sin\theta\sin\phi$. The R.H.S of Eq.~\ref{Eq_boltz4} simplifies to
\begin{eqnarray}
\mathcal{V}\sum_\eta &\beta^{\chi\eta} \iint f^{\eta} (\theta',\phi') \mathcal{G}^{\chi\eta} (\lambda^\eta - h^\eta_\mu(\theta',\phi') + a^\eta \cos\theta' +\nonumber\\
&b^\eta \sin\theta'\cos\phi' + c^\eta \sin\theta'\sin\phi')d\theta'd\phi',
\label{Eq_boltz5_rhs}
\end{eqnarray}
where the function
\begin{eqnarray}
f^{\eta} (\theta',\phi') = \frac{(k')^3}{\mathrm{abs}(\mathbf{v}^\eta_{\mathbf{k}'}\cdot \mathbf{k}'^\eta)} \sin\theta' (\mathcal{D}^\eta_{\mathbf{k}'})^{-1} \tau^\eta_\mu(\theta',\phi')
\label{Eq_f_eta}
\end{eqnarray}
The above equations, when written down explicitly, take the form of seven simultaneous equations to be solved for eight variables. The final constraint comes from charge conservation 
\begin{eqnarray}
\sum\limits_{\eta}\sum\limits_{\mathbf{k}} g^\eta_\mathbf{k} = 0
\label{Eq_sumgk}
\end{eqnarray}
Eq.~\ref{Eq_Lambda_1}, Eq.~\ref{Eq_boltz5_rhs}, Eq.~\ref{Eq_f_eta} and Eq.~\ref{Eq_sumgk} are solved together with Eq~\ref{Eq_tau2}, simultaneously for the eight unknowns ($\lambda^{\pm 1}, a^{\pm 1}, b^{\pm 1}, c^{\pm 1}$). All the two-dimensional integrals w.r.t \{$\theta'$, $\phi'$\}, and the solution of the simultaneous equations are performed numerically. Finally, the current is given by
\begin{align}
    \mathbf{j} = -e\sum\limits_\eta\sum\limits_\mathbf{k} v^\eta_\mathbf{k} f^\eta_\mathbf{k}, 
\end{align}
and $\mathbf{j} = \hat{\sigma} \mathbf{E}$, gives the conductivity, where $\hat{\sigma}$ is the conductivity tensor. The resistivity tensor is given by the inverse of the the conductivity tensor. In figure \ref{fig:theoryplot1} (a), we plot the change in magnetoresistance $\delta\rho_{xx}(B)$ as a function of the magnetic field. Typically the magnetoresistance in Weyl semimetals was thought to be always negative, but beyond a critical intervalley scattering strength, the magnetoresistance switches sign from negative to positive~\cite{sharma2020sign}. We thus ascribe the observed positive AMR to strong intervalley scattering and chiral anomaly in PdTe$_2$. In figure \ref{fig:theoryplot1} (b) we plot the anisotropic magnetoresistance as a function of the angle $\gamma$, which matches well with the experimental data. In Fig.~\ref{fig:theoryplot1} (c) we plot the parametric plots of $\rho_{xy}$ vs. $\rho_{xx}$, which again matches well with the experimental data, As we increase the magnetic field, the center of the plots move to the right because the value of magnetoresistance at $\gamma=\pi/2$ increases (when the fields are parallel to each other). As we vary the angle $\phi$, the planar Hall resistivity starts to acquire a nonzero value. We find that all experimental observations are in good agreement with our theoretical expectations.

\section{Summary}
We have performed the electronic transport (Planar Hall and anisotropic magnetoresistance) and thermal transport (Seebeck and Nernst coefficient) measurements on single crystalline PdTe$_2$, Cu$_{0.05}$PdTe$_2$, and Ag$_{0.05}$PdTe$_2$. Despite the observation of large positive longitudinal MR and prolate pattern in parametric plots, we find that Weyl Physics (chiral anomaly) very well explains all the experimental observations. However, the increase in both $\rho_\perp$ and $\rho_\parallel$ with magnetic field manifest the presence of orbital MR in these systems. The linear field dependence of $\Delta\rho$ suggest intermediate coupling of trivial electric and chiral charge. However, no signature of any coupling between electric and chiral charge has been observed through the temperature dependence of $\Delta\rho$, electrical resistivity and thermal transport data. So, the possibility of Fermi surface reconstruction leading to PHE is ruled out. Therefore, we conclude that the PHE in PdTe$_2$ and its intercalated compounds is governed by chiral anomaly accompanied by orbital MR at high temperatures.
 
A large value of AMR ratio in Ag$_{0.05}$PdTe$_2$ in comparison to many conventional ferromagnets, suggests it to be a potential candidate for future magnetic sensors and magnetic memory devices. The magnitude of both $\it{S}$ and $\nu$ at phonon drag minima increases with increasing magnetic field and in going from PdTe$_2$ to Cu$_{0.05}$PdTe$_2$ to Ag$_{0.05}$PdTe$_2$. Further, intercalation of Cu and Ag results in enhancement of Fermi energy by $\sim$ 2 times and $\sim$ 3 times of PdTe$_2$ respectively.\\

\textbf{Acknowledgment}: We acknowledge Advanced Material Research Center (AMRC), IIT Mandi for the experimental facilities. Sonika and Sunil Gangwar acknowledge IIT Mandi and MHRD India for the HTRA fellowship. We are grateful to Yogesh Singh and Goutam Sheet for the transport measurement in their dilution refrigerator. We acknowledge the Helium plant facility of IISER Mohali. G.S. acknowledges support from SERB Grant No.
SRG/2020/000134 and the IIT Mandi Seed Grant No. IITM/SG/GS/73.
C.S.Y. acknowledges SERB-DST (India) for the CRG grant (CRG/2021/002743).
 
\bibliography{APT}

\section{Supplementary information}

\subsection{Structural analysis (Ag$_{0.05}$PdTe$_2$)}

\begin{figure}[b]
	\includegraphics[width= \columnwidth, height = 12 cm]{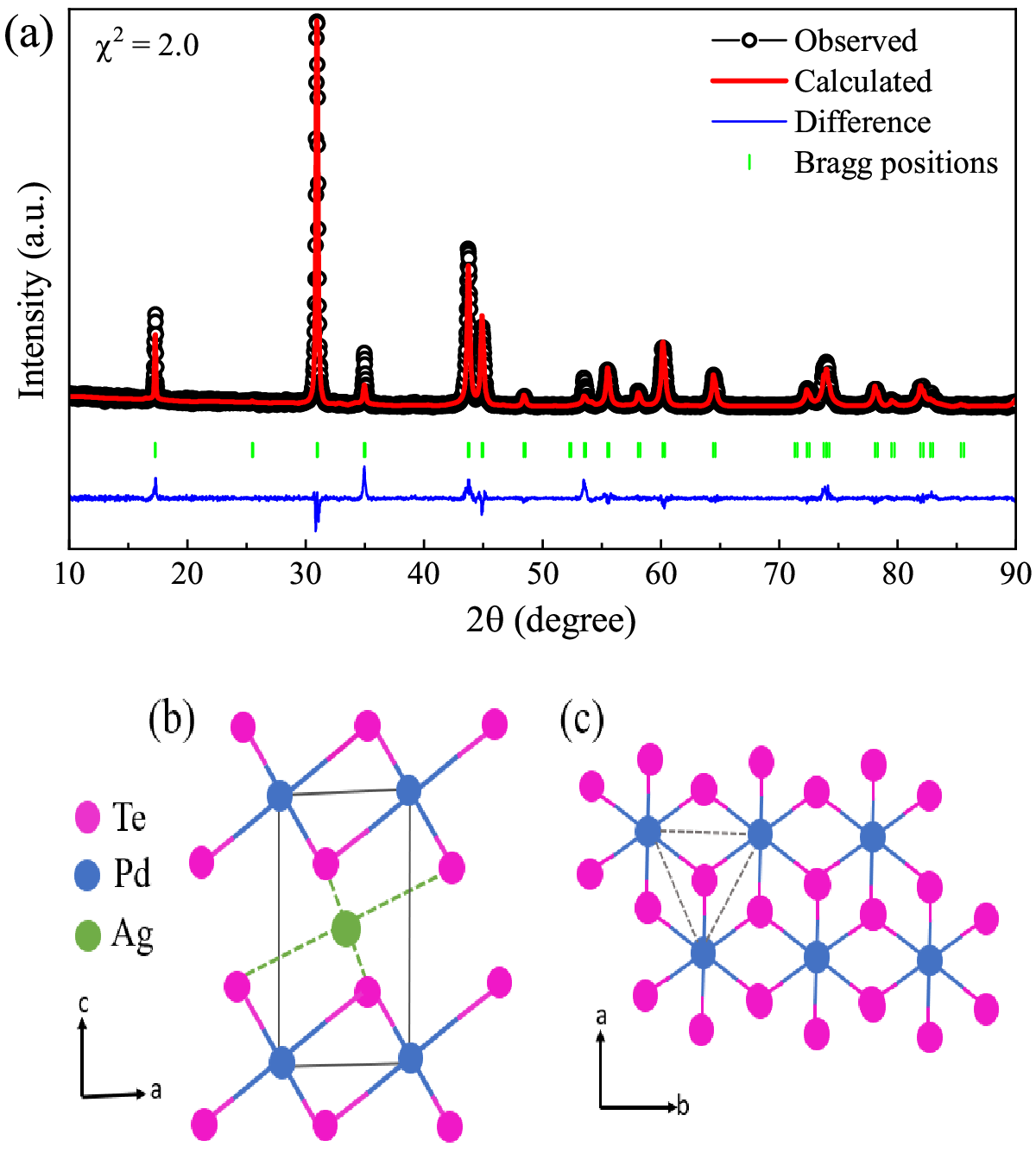}
	\caption{(a) Room temperature Rietveld refined XRD data for powdered Ag$_{0.05}$PdTe$_2$ (b) Side view of crystal structure of CdI$_2$-type PdTe$_2$ doped with Ag (c) Top view of the crystal structure. The Blue, magenta and green spheres indicate Pd, Te and Ag atoms respectively.}
	\label{fig:FigureS1}
\end{figure}

\begin{figure}[t]
	\includegraphics[width= \columnwidth, height = 12 cm]{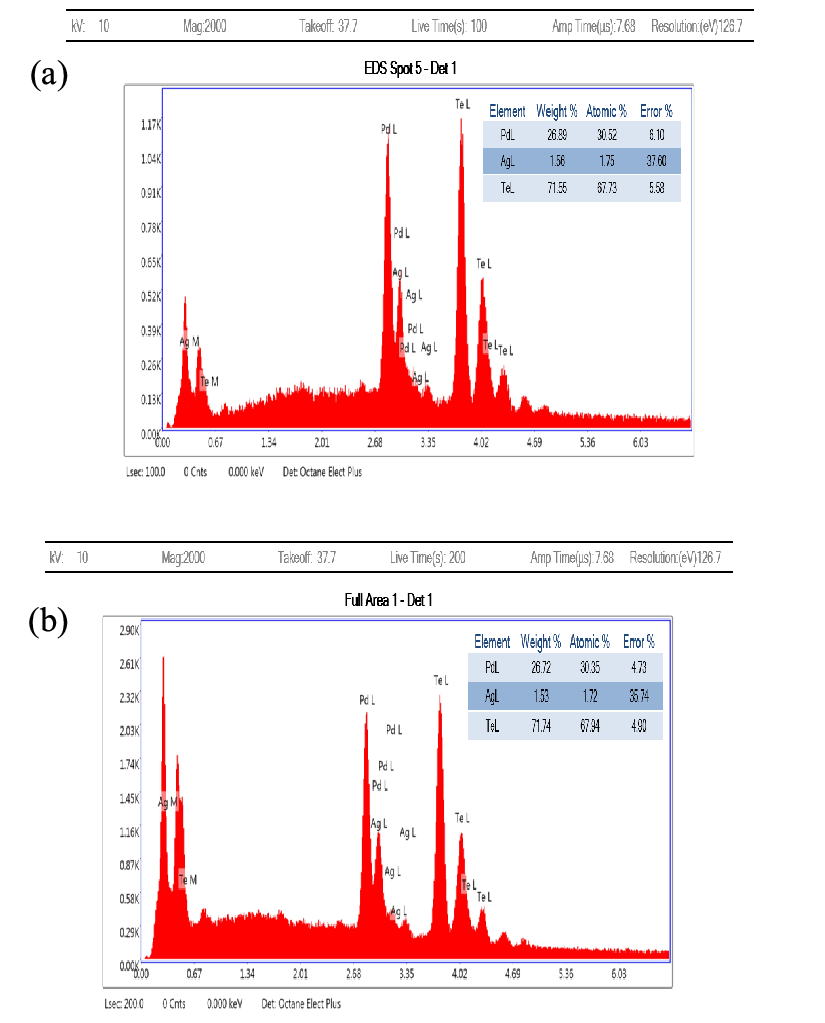}
	\caption{EDX image of Ag doped PdTe$_2$ compound}
	\label{fig:FigureS2}
\end{figure}

\begin{figure}[t]
	\includegraphics[width= \columnwidth, height = 6 cm]{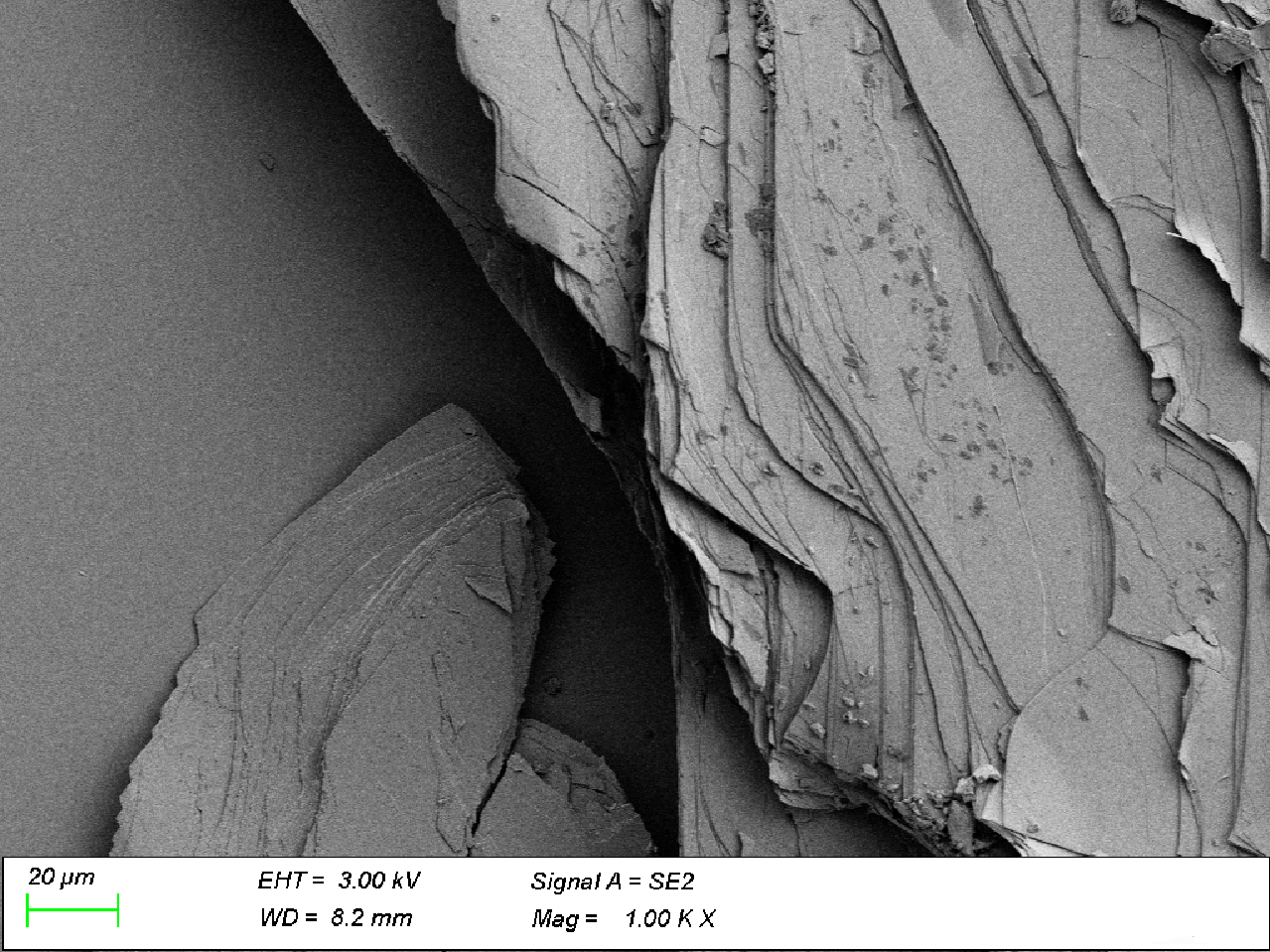}
	\caption{FESEM image of Ag doped PdTe$_2$ compound in the magnification of 1000 times.}
	\label{fig:FigureS3}
\end{figure}

Room temperature powder X-Ray Diffraction (XRD) data for Ag$_{0.05}$PdTe$_2$ has been analysed using Rietveld refinemnet method in FullProf suit software (figure \ref{fig:FigureS1}(a)). The crystal structure for Ag$_{0.05}$PdTe$_2$ is of CdI$_2$-type trigonal structure with space group {\it P}$\bar{3}$m1. The lattice parameters calculated from Rietveld fit are a = b = 4.036 \AA and c = 5.133 \AA. The XRD pattern of the powdered sample indicates formation of sample in single phase. Figure \ref{fig:FigureS1}(b) and \ref{fig:FigureS1}(c) shows the side and top view of the crystal structure of Ag$_{0.05}$PdTe$_2$ where Pd and Te atoms have coordination numbers of six and three respectively. In this layered compound, there is van der Waals gap between the Te anion layers which allows the effective intercalation of cation such as Ag or Cu that has a monovalent valence state, within the layers. As shown in figure \ref{fig:FigureS1}(c), the center position of triangular geometry in the trigonal phase is the most ideal site for cation intercalation.

The elemental composition of compound calculated from Energy dispersive X-Ray spectroscopy (EDX) averaged over five different spots two of which are shown in figure \ref{fig:FigureS2} is Ag$_{0.04}$PdTe$_{2.25}$. This indicate that the amount of Ag contained is less than the nominal value taken during sample preparation. Figure \ref{fig:FigureS3} shows the field emission scanning electron microscopy (FESEM) image of Ag intercalted PdTe$_2$ compound. The image shows the layered morphology of the sample. 

\subsection{Magnetoresistance (PdTe$_2$)} 

\begin{figure}[tb]
	\includegraphics[width= \columnwidth, height = 7 cm]{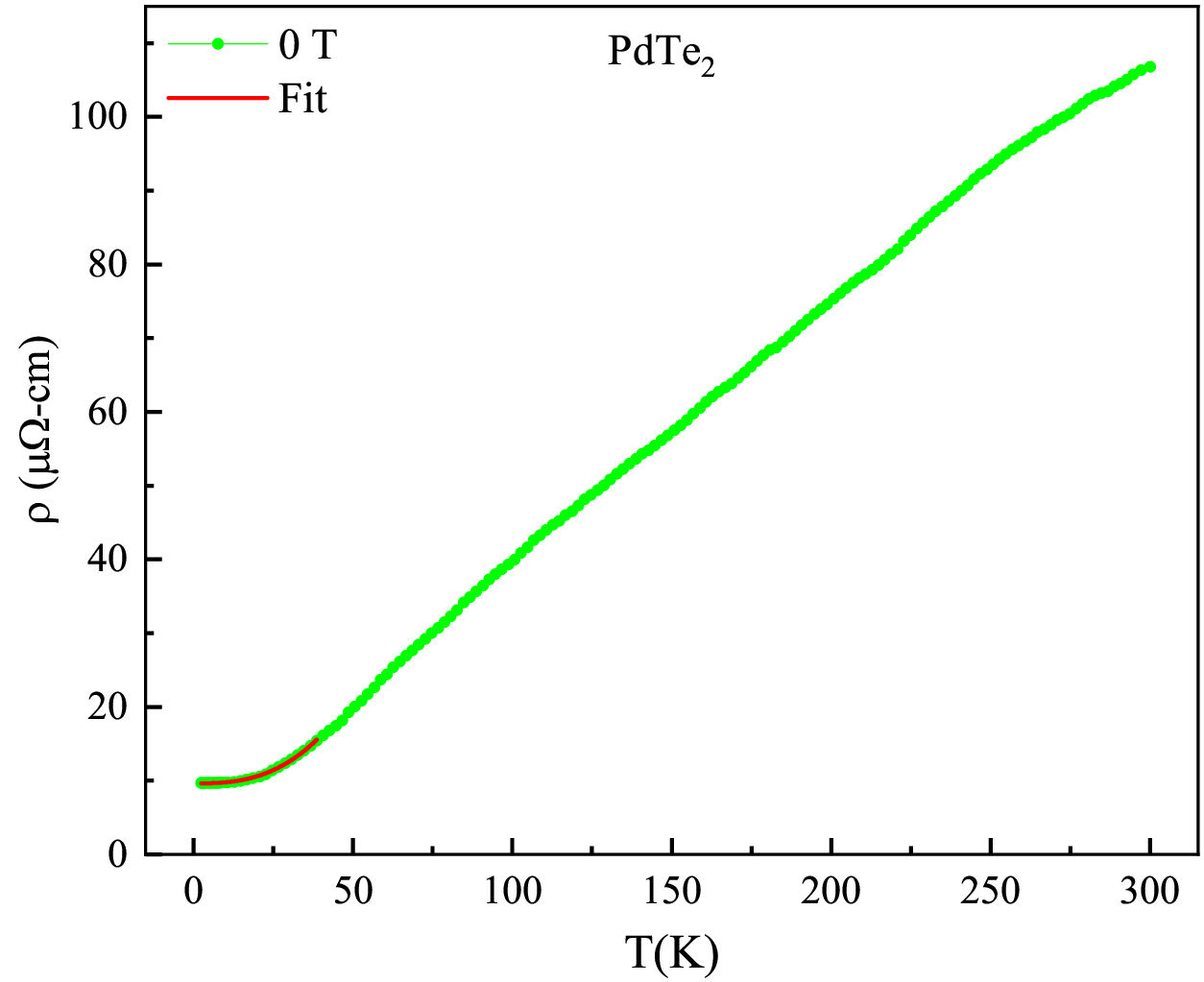}
	\caption{Temperature dependence of electrical resistivity of PdTe$_2$ at $\it{\mu_0 H}$ = 0 Tesla. The red curve shows $\it{T^3}$ dependence for \textit{T} = 2 - 30 K}
	\label{fig:FigureS4}
\end{figure}

\begin{figure}[tb]
	\includegraphics[width= \columnwidth, height = 7 cm]{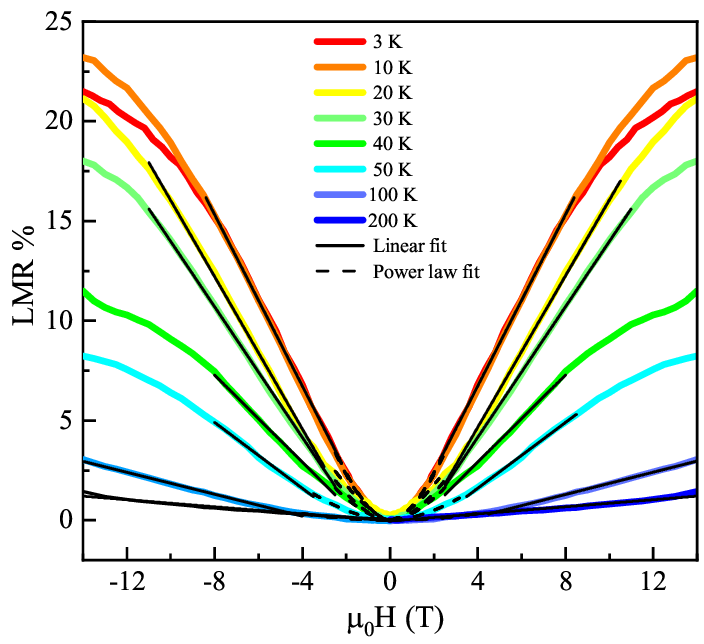}
	\caption{Longitudinal MR for PdTe$_2$ at different temperatures}
	\label{fig:FigureS5}
\end{figure}

The temperature dependence of electrical resistivity ($\rho$(T)) of PdTe$_2$ in the absence of magnetic field ($\it{\mu_0 H}$) is shown in figure \ref{fig:FigureS4}. The ($\rho$(T)) for PdTe$_2$ shows a $\it{T}^3$ dependence for $\it{T}$ = 2 - 30 K just like that of Ag$_{0.05}$PdTe$_2$. The longitudinal magnetoresistance (MR = ($\rho$($\it{\mu_0 H}$) - $\rho$(0))/$\rho$(0)) for $\it{\mu_0 H} \parallel \it{I}$ is measured for \textit{T} range 3 - 200 K for $\it{\mu_0 H}$ =  0 - 14 Tesla is shown in figure \ref{fig:FigureS5} for PdTe$_2$. The MR is positive for the measured temperature range and the value of MR increases on lowering \textit{T}. The MR shows linear field dependence at high \textit{T} (100 - 200 K) for the full $\it{\mu_0 H}$ range of $\it{\mu_0 H}$ = 0 - 14 Tesla. For the low \textit{T} = 3 - 50 K, MR show linear $\it{\mu_0 H}$ dependence for $\it{\mu_0 H} >$ 3 Tesla and power law field dependence ($\sim$ a$\it{(\mu_0 H)^m}$; where a is a proportionality constant and \textit{m} = 1.5 - 1.7 is the power index) for $\it{\mu_0 H} <$ 3 Tesla. 

\subsection{Planar Hall effect and Anisotropic Magnetoresistance (PdTe$_2$)}

\begin{figure}[tb]
	\includegraphics[width= \columnwidth, height = 10 cm]{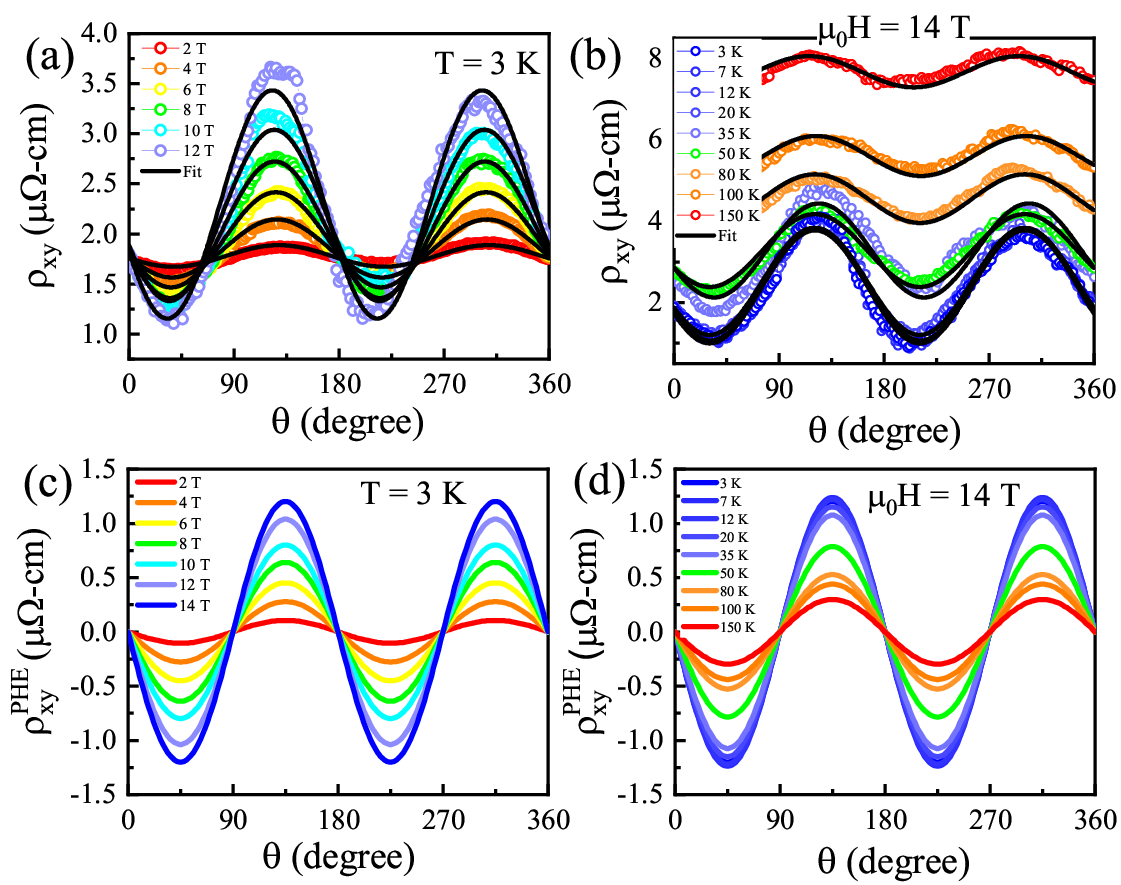}
	\caption{(a) Angular dependence of $\rho_{xy}$ measured in planar Hall configuration at different $\it{\mu_0 H}$ for \textit{T} = 3 K and (b) at different \textit{T} for $\it{\mu_0 H}$ = 14 Tesla for PdTe$_2$. Fitting of data is shown by black curves. Extracted value of planar Hall signal (c) at different $\it{\mu_0 H}$ for \textit{T} = 3 K and (d) at different \textit{T} for $\it{\mu_0 H}$ = 14 Tesla.}
	\label{fig:FigureS6}
\end{figure}

\begin{figure}[tb]
	\includegraphics[width= \columnwidth, height = 10 cm]{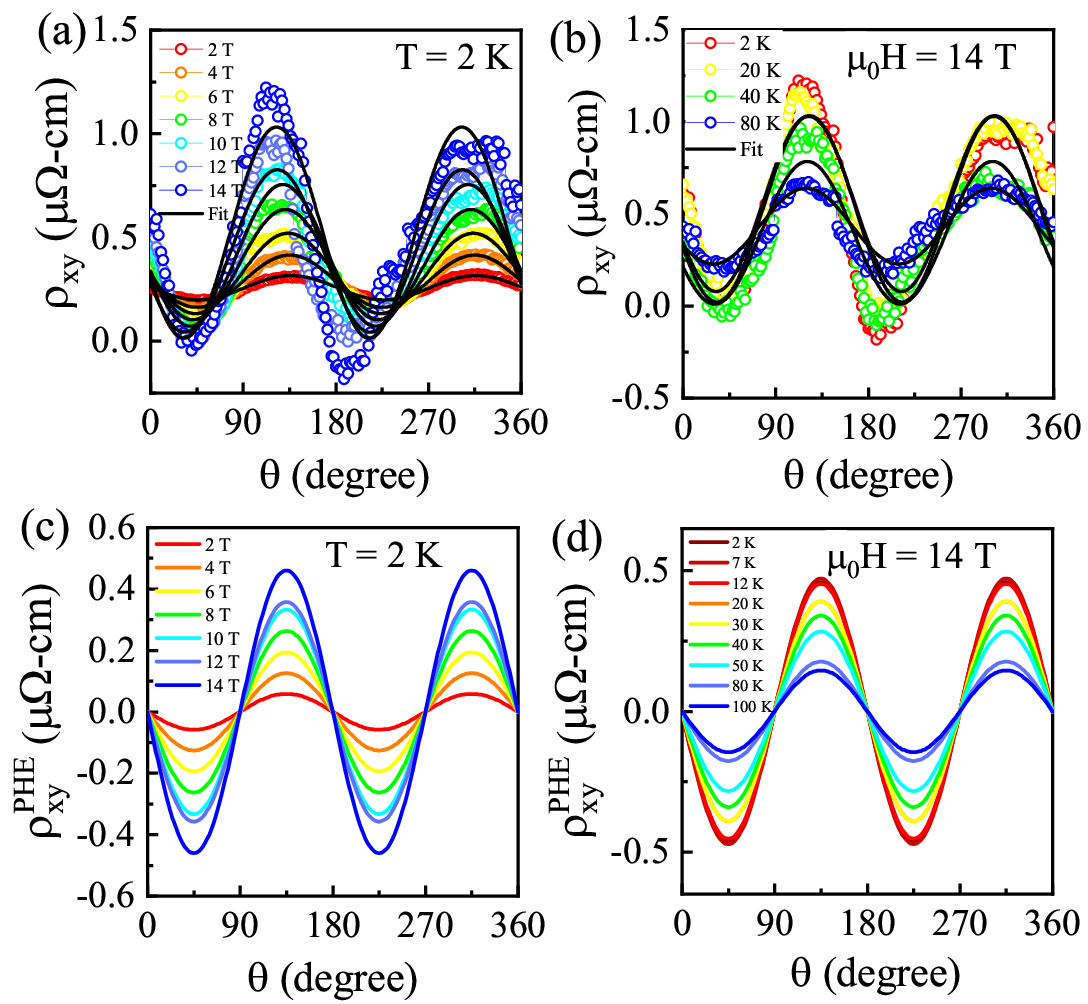}
	\caption{(a) Angular dependence of $\rho_{xy}$ measured in planar Hall configuration at different $\it{\mu_0 H}$ for \textit{T} = 2 K and (b) at different \textit{T} for $\it{\mu_0 H}$ = 14 Tesla for Cu$_{0.05}$PdTe$_2$. Fitting of data is shown by black curves. Extracted value of planar Hall signal (c) at different $\it{\mu_0 H}$ for \textit{T} = 2 K and (d) at different \textit{T} for $\it{\mu_0 H}$ = 14 Tesla.}
	\label{fig:FigureS7}
\end{figure}

\begin{figure}
	\includegraphics[width= 8 cm, height = 10 cm]{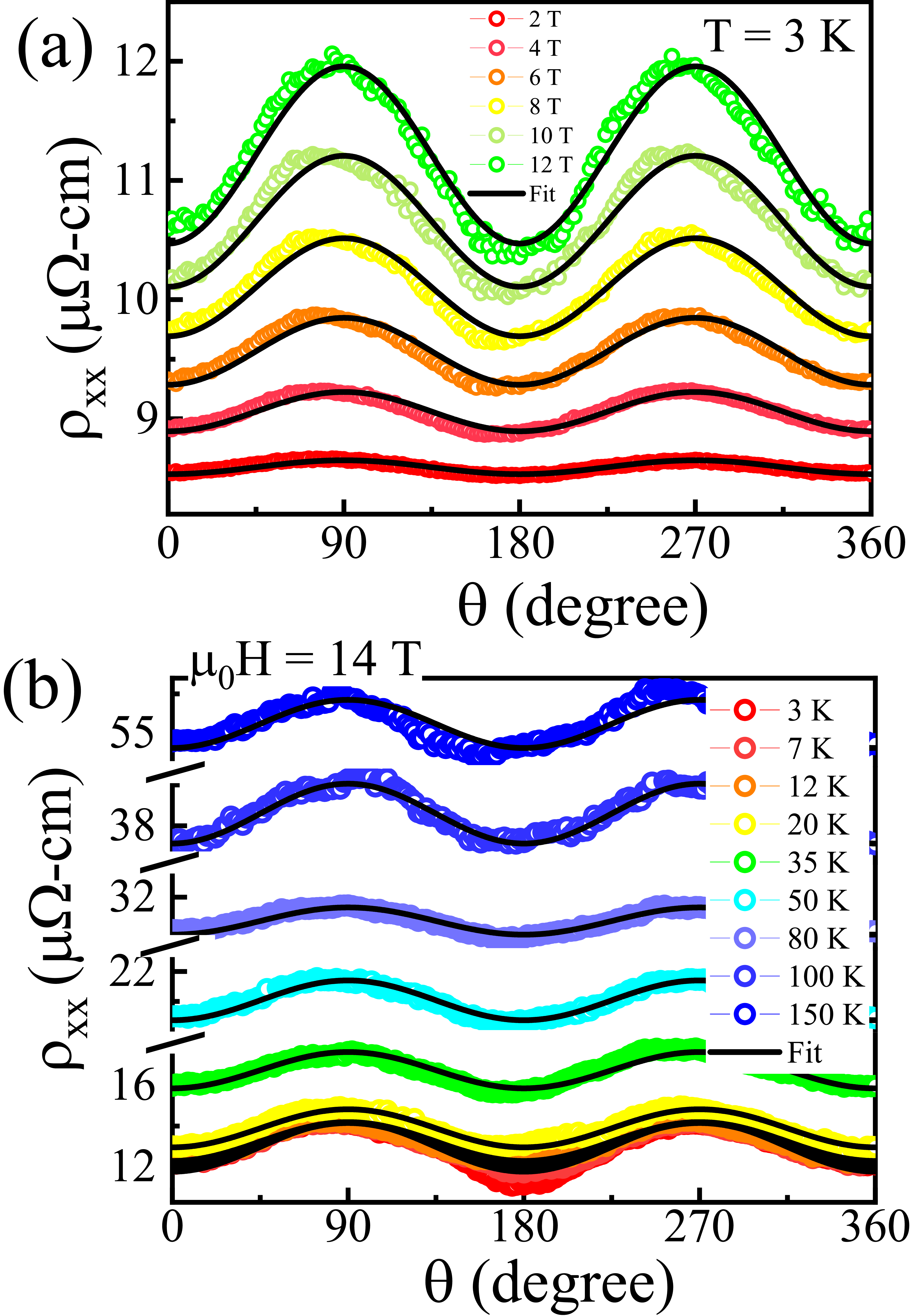}
	\caption{Angular dependence of $\rho_{xx}$ for PdTe$_2$ (a) at different $\it{\mu_0 H}$ for \textit{T} = 3 K and (b) at different \textit{T} for $\it{\mu_0 H}$ = 14 Tesla.}
	\label{fig:FigureS8}
\end{figure}

\begin{figure}
	\includegraphics[width= 8 cm, height = 10 cm]{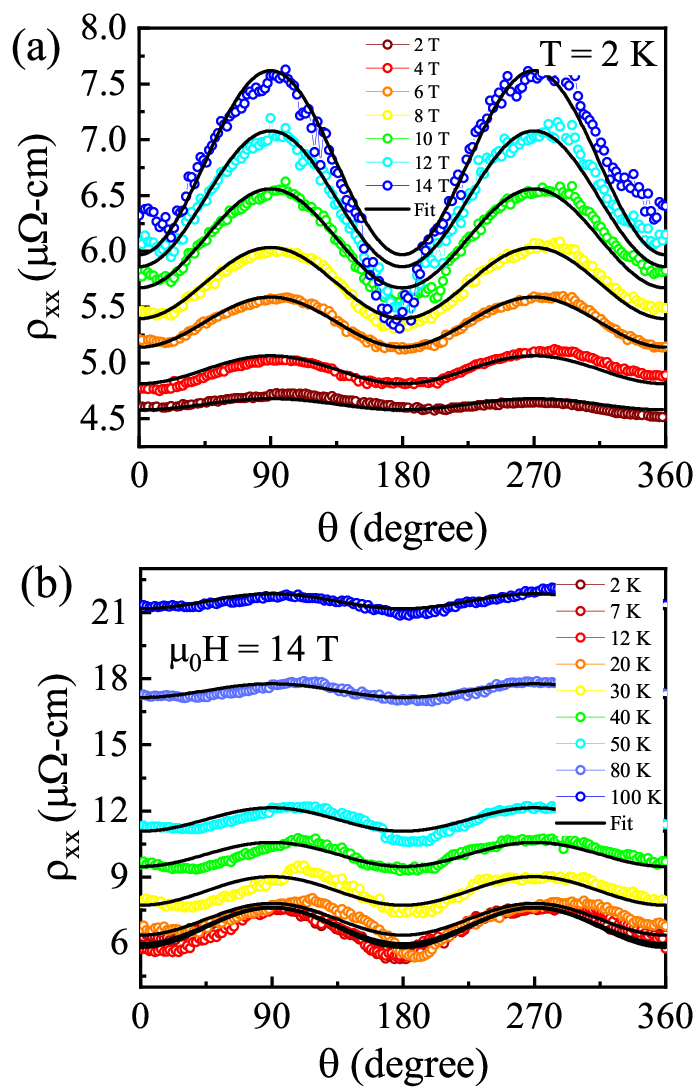}
	\caption{Angular dependence of $\rho_{xx}$ for Cu$_{0.05}$PdTe$_2$ (a) at different $\it{\mu_0 H}$ for \textit{T} = 2 K and (b) at different \textit{T} for $\it{\mu_0 H}$ = 14 Tesla.}
	\label{fig:FigureS9}
\end{figure}

The angular dependence of transverse resistivity ($\rho_{xy}$) for PdTe$_2$ at different $\it{\mu_0 H}$ for $\it{T}$ = 3 K and at different \textit{T} for $\it{\mu_0 H}$ = 14 Tesla is shown in figure \ref{fig:FigureS6}(a) and \ref{fig:FigureS6}(b) respectively. The planar Hall resistivity has been extracted from the fitting of $\rho_{xy}$ using following equation \cite{nandy2017chiral, burkov2017giant}.

\begin{equation}
	\rho_{xy} = \rho^{PHE}_{xy} + \rho_{long.} + \rho_{G}
	\label{eq1}
\end{equation}

where $\rho^{PHE}_{xy} = -\Delta\rho sin\theta cos\theta$, $\rho_{long.} = \Delta\rho cos^2\theta$ and $\Delta\rho$ = $\rho_{\perp}$ - $\rho_{\parallel}$ is the amplitude of PHE. $\rho_{\parallel}$ and $\rho_{\perp}$ are the resistivities corresponding to $\it{\mu_0 H} \parallel \textit{I}$ and $\it{\mu_0 H} \perp \textit{I}$ respectively.

The angular dependence of planar Hall resistivity at different $\it{\mu_0 H}$ for $\it{T}$ = 3 K and at different \textit{T} for $\it{\mu_0 H}$ = 14 Tesla is shown in figure \ref{fig:FigureS6}(c) and \ref{fig:FigureS6}(d) respectively. Similar, angular dependence has been observed for Cu$_{0.05}$PdTe$_2$ for $\it{T}$ = 2 K and $\it{\mu_0 H}$ = 14 Tesla. The planar Hall resistivity for PdTe$_2$ and Cu$_{0.05}$PdTe$_2$ follows the trend similar to that of Ag$_{0.05}$PdTe$_2$ with reduced values at same \textit{T} and $\it{\mu_0 H}$ (figure \ref{fig:FigureS7}).

Figure \ref{fig:FigureS8}(a) and \ref{fig:FigureS8}(b) shows AMR at different $\it{\mu_0 H}$ for $\it{T}$ = 3 K and at different \textit{T} for $\it{\mu_0 H}$ = 14 Tesla respectively for PdTe$_2$. The AMR data oscillates with a period of $\pi$ with minima and maxima at ${0}^o$ and ${90}^o$ respectively as per following equation \cite{nandy2017chiral, burkov2017giant} 

\begin{equation}
	\rho_{xx} = \rho_{\perp} - \Delta\rho cos^2\theta
	\label{eq2}
\end{equation}

AMR data is fitted using equation \ref{eq2} and is used to extract the values of $\rho_{\parallel}$ and $\rho_{\perp}$. The AMR ratio for PdTe$_2$ is $\sim$ -12 at $\it{\mu_0 H}$ = 12 Tesla and $\it{T}$ = 3 K, which is less than that observed for Ag$_{0.05}$PdTe$_2$. The angular dependence of AMR at different $\it{\mu_0 H}$ for $\it{T}$ = 2 K and at different \textit{T} for $\it{\mu_0 H}$ = 14 Tesla respectively for Cu$_{0.05}$PdTe$_2$ is shown in figure \ref{fig:FigureS9}(a) and \ref{fig:FigureS9}(b). 

\subsection{Thermal transport}

\begin{figure*}
	\includegraphics[height = 10 cm]{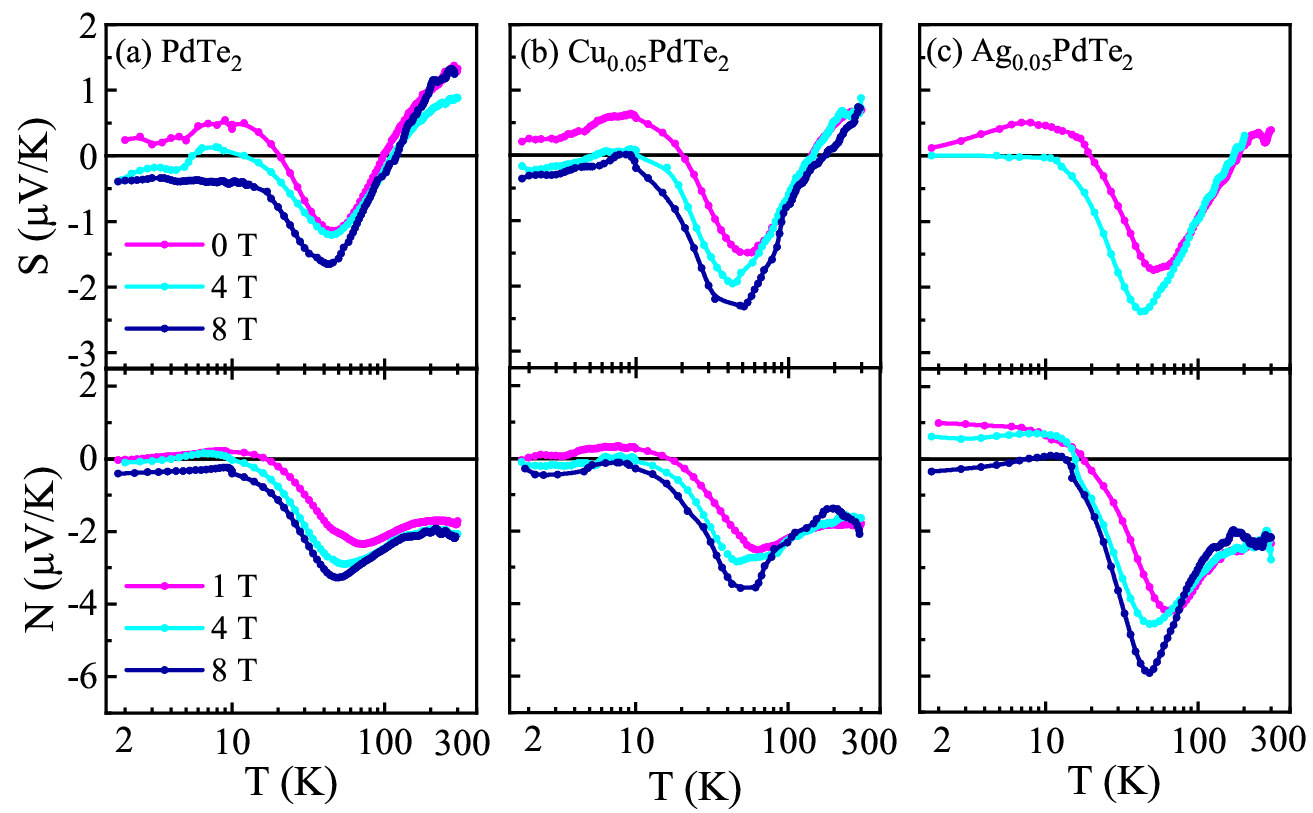}
	\caption{Seebeck coefficient (upper panel) and Nernst coefficient (lower panel) of (a) PdTe$_2$ (b) Cu$_{0.05}$PdTe$_2$ and (c) Ag$_{0.05}$PdTe$_2$ at different magnetic fields.}
	\label{fig:FigureS10}
\end{figure*}

The Seebeck ($\it{S}$) and Nernst ($\it{N}$) coefficient  of three compounds PdTe$_2$, Cu$_{0.05}$PdTe$_2$, and Ag$_{0.05}$PdTe$_2$ measured for $\it{T}$ = 1.8 - 300 K is shown in the upper and lower panel of figure \ref{fig:FigureS8} (a-c) respectively. The positive to negative crossover in Seebeck data for the three compounds PdTe$_2$ (99 K), Cu$_{0.05}$PdTe$_2$ (144 K), and Ag$_{0.05}$PdTe$_2$ (183 K) increases on increasing the magnetic field. The Nernst coefficient is negative for $\it{T}$ = 18 - 300 K for three compounds. The negative to positive crossover temperature in $\it{N}$ decreases with the increase in magnetic field in all three samples. The phonon drag minima for Seebeck coefficient of three compounds PdTe$_2$, Cu$_{0.05}$PdTe$_2$, and Ag$_{0.05}$PdTe$_2$ occurs at 45 K, 54 K and 51 K respectively. However, the phonon drag minima in the Nernst coefficient of three compounds PdTe$_2$, Cu$_{0.05}$PdTe$_2$, and Ag$_{0.05}$PdTe$_2$ occurs at 70 K, 63 K and 68 K respectively. These minima shifts towards the lower temperatures with the increase in magnetic field in both $\it{S}$ and $\it{N}$. Also, it has been observed that the magnitude of $\it{S}$ and $\it{N}$ at phonon drag minima increases with increasing magnetic field and in going from PdTe$_2$ to Cu$_{0.05}$PdTe$_2$ to Ag$_{0.05}$PdTe$_2$. The new peak observed at $\sim{9}$ K is also observed to shift towards lower temperature with increase in magnetic field in both $\it{S}$ and $\it{N}$.

\end{document}